%% file: SUS-20-003_temp.tex
\pdfoutput=1
\documentclass[11pt,twoside,a4paper,cmspaper,final,collab]{cms-tdr}
\def\svnVersion{899304f-D}\def\svnDate{2021/08/20}\def\cmsCernNoTag{CERN-EP-2021-114}\def\cmsCernDate{\today}\def\cmsMessage{Published in the Journal of High Energy Physics as \href{http://dx.doi.org/10.1007/JHEP10(2021)045}{\doi{10.1007/JHEP10(2021)045}.}}
\begin{document}\cmsNoteHeader{SUS-20-003}

\newcommand{\mbb}{\ensuremath{m_{\bbbar}}\xspace}
\newcommand{\wjets}{\ensuremath{\PW{}\text{+jets}}\xspace}
\newcommand{\whfjets}{\ensuremath{\PW{}\mathrm{+HF}}\xspace}
\newcommand{\dyhfjets}{\ensuremath{\mathrm{DY}\mathrm{+HF}}\xspace}
\newcommand{\wlfjets}{\ensuremath{\PW{}\text{+LF}}\xspace}
\newcommand{\ttjets}{\ensuremath{\ttbar{}\text{+jets}}\xspace}
\newcommand{\Lint}{\ensuremath{137\fbinv}\xspace}
\newcommand{\mct}{\ensuremath{m_{\mathrm{CT}}}\xspace}
\newcommand{\rw}{\ensuremath{R_{\PW}}\xspace}
\newcommand{\rmct}{\ensuremath{R_{\text{top}}}\xspace}
\newlength\cmsTabSkip\setlength{\cmsTabSkip}{1ex}
\newcommand{\Nb}{\ensuremath{N_{\PQb}}\xspace}
\newcommand{\ptsum}{\ensuremath{\pt^{\text{sum}}}\xspace}
\newcommand{\pp}{\ensuremath{\Pp{}\Pp{}}\xspace}
\newcommand{\njets}{\ensuremath{N_{\text{jets}}}\xspace}
\newcommand{\nhiggs}{\ensuremath{N_{\PH}}\xspace}
\newcommand{\nh}{\ensuremath{N_{\PH}}\xspace}
\newcommand{\nH}{\ensuremath{N_{\PH}}\xspace}
\newcommand{\CR}{\ensuremath{\mathrm{CR}}\xspace}
\newcommand{\SR}{\ensuremath{\mathrm{SR}}\xspace}
\newcommand{\MCT}{\ensuremath{m_{\mathrm{CT}}}\xspace}

\cmsNoteHeader{SUS-20-003} 
\title{Search for chargino-neutralino production in events with Higgs and \texorpdfstring{\PW}{W} bosons using 137\fbinv of proton-proton collisions at \texorpdfstring{$\sqrt{s}=13\TeV$}{sqrt(s) = 13 TeV}}

\date{\today}

\abstract{
A search for electroweak production of supersymmetric (SUSY) particles in final states with one lepton, a Higgs boson decaying to a pair of bottom quarks, and large missing transverse momentum is presented. The search uses data from proton-proton collisions at a center-of-mass energy of 13\TeV collected using the CMS detector at the LHC, corresponding to an integrated luminosity of 137\fbinv. The observed yields are consistent with backgrounds expected from the standard model. The results are interpreted in the context of a simplified SUSY model of chargino-neutralino production, with the chargino decaying to a \PW boson and the lightest SUSY particle (LSP) and the neutralino decaying to a Higgs boson and the LSP. Charginos and neutralinos with masses up to 820\GeV are excluded at $95\%$ confidence level when the LSP mass is small, and LSPs with mass up to 350\GeV are excluded when the masses of the chargino and neutralino are approximately 700\GeV.}

\hypersetup{%
pdfauthor={CMS Collaboration},%
pdftitle={Search for chargino-neutralino production in events with Higgs and W bosons using 137 fb-1 of proton-proton collisions at sqrt(s) = 13 TeV},%
pdfsubject={CMS},%
pdfkeywords={CMS,  supersymmetry, boosted, electroweakino, Higgs, chargino, neutralino}}

\maketitle 

\section{Introduction}
\label{sec:intro}

Supersymmetry (SUSY)~\cite{Ramond:1971gb,Wess:1974tw,Nilles:1983ge} is an appealing extension of the standard model (SM) that predicts the existence of a superpartner for every SM particle, with the same gauge quantum numbers but differing by one half unit of spin. SUSY allows addressing several shortcomings of the SM. For example, the superpartners can play an important role in stabilizing the mass of the Higgs boson (\PH{})~\cite{deCarlos:1993rbr,Dimopoulos:1995mi}. In \textit{R}-parity conserving SUSY models, the lightest supersymmetric particle (LSP) is stable and therefore is a viable dark matter candidate~\cite{Farrar:1978xj}.

The SUSY partners of the SM gauge bosons and the Higgs boson are known as winos (partners of the SU(2)$_{\mathrm{L}}$ gauge fields), the bino (partner of the U(1) gauge field), and higgsinos.
Neutralinos (\PSGcz) and charginos (\PSGcpm) are the corresponding mass eigenstates of the winos, bino and higgsinos.
They do not carry color charge and are therefore produced only via electroweak interactions or in the decay of colored superpartners. Because of the smaller cross sections for electroweak processes, the masses of these particles are experimentally less constrained than the masses of colored SUSY particles. Depending on the mass spectrum, the neutralinos and charginos can have significant decay branching fractions to vector or scalar bosons. In particular, the decays via the \PW and the Higgs boson are expected to be significant if the \PSGcpmDo and \PSGczDt particles are wino-like, the \PSGczDo is bino-like, and the difference between their masses is larger than the Higgs boson mass, where the subscript $1(2)$ denotes the lightest (second lightest) neutralino or chargino, respectively. These considerations strongly motivate a search for the electroweak production of SUSY partners presented in this paper. 

\begin{figure}[bh]
\centering
\includegraphics[width=0.45 \textwidth]{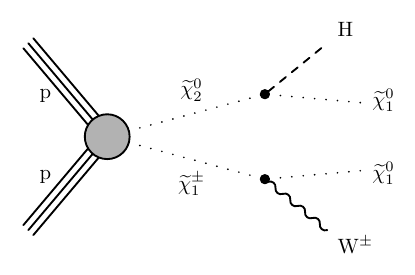}
\caption{
Diagram for a simplified SUSY model with electroweak production of the lightest chargino \PSGcpmDo and next-to-lightest neutralino \PSGczDt. The \PSGcpmDo decays to a \PW boson and the lightest neutralino \PSGczDo. The \PSGczDt decays to a Higgs boson and a \PSGczDo.}
\label{fig:fey}
\end{figure}

This paper reports the results of a search for chargino-neutralino production with subsequent $\PSGcpmDo \to \PW^{\pm} \PSGczDo$ and $\PSGczDt \to \PH \PSGczDo$ decays, as shown in Fig.~\ref{fig:fey}. The data analysis focuses on the final state with a charged lepton produced in the \PW boson decay, two jets reconstructed from the $\PH \to \PQb\PAQb$ decay, and significant missing transverse momentum (\ptmiss) resulting from the LSPs and the neutrino. This final state benefits from the large branching fraction for $\PH \to \PQb\PAQb$, 58\%. The chargino and neutralino are assumed to be wino-like, and the \PSGczDo produced in their decays is assumed to be the stable LSP. As wino-like charginos $\PSGcpmDo$ and neutralinos $\PSGczDt$ would be nearly degenerate, this analysis considers a simplified model~\cite{Alwall:2008ag,Alves:2011wf,Chatrchyan:2013sza} with a single mass parameter for both the chargino and neutralino ($m_{\PSGczDt/\PSGcpmDo}$), as well as an additional mass parameter for the LSP ($m_{\PSGczDo}$). Results of searches in this final state were previously presented by ATLAS~\cite{ATLAS_WH_run2,Aad:2015eda} and CMS~\cite{Sirunyan:2017zss,Khachatryan:2014mma,Khachatryan:2014qwa} using data sets at center of mass energy 8 and 13\TeV.

This analysis uses 13\TeV proton-proton (\pp) collision data collected with the CMS detector during the 2016--2018 data-taking periods, corresponding to an integrated luminosity of \Lint. Relative to the most recent result from the CMS Collaboration targeting this signature~\cite{Sirunyan:2017zss}, the results significantly extend the sensitivity to the mass of the chargino and neutralino. The improved sensitivity is achieved through a nearly four-fold increase in the integrated luminosity, as well as from numerous improvements in the analysis, including the addition of a discriminant that identifies Higgs boson decays collimated into large-radius jets, regions that include additional jets from the initial-state radiation, and an expanded categorization in \ptmiss.

\section{The CMS detector}
\label{sec:CMS}

The central feature of the CMS apparatus is a superconducting solenoid of 6\unit{m} internal diameter, providing a magnetic field of 3.8\unit{T}. Within the solenoid volume are a silicon pixel and strip tracker, a lead tungstate crystal electromagnetic calorimeter, and a brass and scintillator hadron calorimeter, each composed of a barrel and two endcap sections. Forward calorimeters extend the pseudorapidity ($\eta$) coverage provided by the barrel and endcap detectors. Muons are detected in gas-ionization chambers embedded in the steel flux-return yoke outside the solenoid.  A more detailed description of the CMS detector, together with a definition of the coordinate system used and the relevant kinematic variables, can be found in Ref.~\cite{Chatrchyan:2008zzk}.

Events of interest are selected using a two-tiered trigger system. The first level, composed of custom hardware processors, uses information from the calorimeters and muon detectors to select events at a rate of approximately 100\unit{kHz} within a fixed time interval of about 4\mus~\cite{Sirunyan:2020zal}. The second level, known as the high-level trigger, consists of a farm of processors running a version of the full event reconstruction software optimized for fast processing, and reduces the event rate to approximately 1\unit{kHz} before data storage~\cite{Khachatryan:2016bia}.

\section{Simulated samples}
\label{sec:mc}

\tolerance=800
Monte Carlo (MC) simulation is used to design the search strategy, study and estimate SM backgrounds, and evaluate the sensitivity of the analysis to the SUSY signal. 
Separate MC simulations are used to reflect the detector configuration and running conditions of different periods (2016, 2017, and 2018).
The \MGvATNLO~2 (versions 2.2.2 for 2016 and 2.4.2 for 2017--2018) generator~\cite{Alwall:2014hca} at leading order (LO) in quantum chromodynamics (QCD) is used to generate samples of events of SM $\ttbar$, $\wjets$, and $\PW\PH$ processes, as well as chargino-neutralino production, as described by a simplified model of SUSY. 
Samples of $\wjets$, \ttbar, and SUSY events are generated with up to four, three, and two additional partons included in the matrix-element calculations, respectively.
The \MGvATNLO generator at next-to-leading-order (NLO) in QCD is used to generate samples of $\ttbar\PZ$ and $\PW\PZ$ events, while single top quark events are generated at NLO in QCD using the \POWHEG~2.0~\cite{Nason:2004rx,Frixione:2007vw,Alioli:2010xd,Re:2010bp} program. 
\par

The NNPDF3.0 (3.1) parton distribution functions, PDFs, are used to generate all 2016 (2017--2018) MC samples~\cite{Ball:2011uy,Ball:2014uwa,Ball:2017nwa}.
The parton shower and hadronization are modeled with \PYTHIA~8.226~(8.230)~\cite{Sjostrand:2014zea} in 2016 (2017--2018) samples.
The MLM~\cite{Alwall:2007fs} and FxFx~\cite{Frederix:2012ps} prescriptions are employed to match partons from the matrix-element calculation to those from the parton showers for the LO and NLO samples, respectively.

The 2016 MC samples are generated with the \textsc{CUETP8M1} \PYTHIA\ tune~\cite{Khachatryan:2015pea}.
For later data-taking periods, the \textsc{CP5} and \textsc{CP2} tunes~\cite{Sirunyan:2019dfx} are used
for SM and SUSY signal samples, respectively.
The \GEANTfour~\cite{Agostinelli2003250} package is used to simulate the response of the CMS detector for all SM processes, while the CMS fast simulation program~\cite{Abdullin:2011zz,Giammanco:2014bza} is used for signal samples.

Cross section calculations performed at next-to-next-to-leading-order (NNLO) in QCD are used to normalize
the MC samples of $\wjets$~\cite{Li:2012wna}, and at NLO in QCD to normalize single top quark samples~\cite{Aliev:2010zk,Kant:2014oha}.
The $\ttbar$ samples are normalized to a cross section determined at NNLO in QCD that includes the resummation of the next-to-next-to-leading-logarithmic soft-gluon terms~\cite{Beneke:2011mq,Cacciari:2011hy,Czakon:2011xx,Baernreuther:2012ws,Czakon:2012zr,Czakon:2012pz,Czakon:2013goa}.
MC samples of other SM background processes are normalized to cross sections obtained from the MC event generators at either LO or NLO in QCD.
Cross sections for wino-like chargino-neutralino production are computed at approximate NLO plus next-to-leading logarithmic (NLL) precision.
Other SUSY particles except for the LSP are assumed to be heavy and decoupled~\cite{PhysRevLett.83.3780,DEBOVE201151,resummino,PhysRevD.98.055014}. A SM-like $\PH \to \PQb \PAQb$ branching fraction of 58.24\%~\cite{deFlorian:2016spz} is assumed.

Nominal distributions of additional \pp collisions in the same or adjacent bunch crossings (pileup) are used in the generation of simulated samples.
These samples are reweighted such that the number of interactions per bunch crossing matches the observation.

\section{Event selection and search strategy}
\label{sec:evtsel}

In order to search for the chargino-neutralino production mechanism shown in Fig.~\ref{fig:fey}, the analysis targets decay modes of the \PW boson to leptons and the \PH to a bottom quark-antiquark pair.
The analysis considers events with a single isolated electron or muon, two jets identified as originating from two bottom quarks, and large \ptmiss from the LSPs and the neutrino.
The major backgrounds in this final state arise from SM processes containing top quarks and \PW bosons.
These backgrounds are suppressed with the analysis strategy described below that uses physics objects summarized in Table~\ref{tab:objs}, which are similar to those presented in Ref.~\cite{Sirunyan_2020_032}.

Events are reconstructed using the particle-flow (PF) algorithm~\cite{Sirunyan:2017ulk}, which combines information from the CMS subdetectors to identify charged and neutral hadrons, photons, electrons, and muons, collectively referred to as PF candidates.
These candidates are associated with reconstructed vertices, and the vertex with the largest sum of squared physics-object transverse momenta is taken to be the primary \pp interaction vertex.
The physics objects used for the primary vertex determination include a special collection of jets reconstructed by clustering only tracks associated to the vertex, and the magnitude of the associated missing transverse momentum. The missing transverse momentum in this case is defined as the negative vector sum of the transverse momentum (\pt) of the jets in this collection. In all other cases, the missing transverse momentum (\ptvecmiss) is taken as the negative vector sum of the \pt of all PF candidates, excluding charged hadron candidates that do not originate from the primary vertex~\cite{Sirunyan:2019kia}.

Electron candidates are reconstructed by combining clusters of energy deposits in the electromagnetic calorimeter with charged tracks~\cite{Sirunyan:2020ycc}.
The electron identification is performed using shower shape variables, track-cluster matching variables, and track quality variables.
The selection on these variables is optimized to identify electrons from the decay of \PW and \PZ bosons while rejecting electron candidates originating from jets.
To reject electrons originating from photon conversions inside the detector, electrons are required to have at most one missing measurement in the innermost tracker layers and to be incompatible with any conversion-like secondary vertices.
Muon candidates are reconstructed by geometrically matching tracks from measurements in the muon system and tracker, and fitting them to form a global muon track.
Muons are selected using the quality of the geometrical matching and the quality of the tracks~\cite{Sirunyan:2018fpa}.

Selected muons (electrons) are required to have $\pt > 25\: (30)\GeV$, $\abs{\eta} < 2.1\: (1.44)$, and be isolated.
Events containing electrons with $\abs{\eta} > 1.44$ have been found to exhibit an anomalous tail in the transverse mass distribution and are not included in the search. 
Lepton isolation is determined from the scalar \pt sum ($\ptsum$) of PF candidates not associated with the lepton within a cone of \pt-dependent radius starting at $\Delta R = \sqrt{\smash[b]{\left(\Delta\phi\right)^{2}+\left(\Delta\eta\right)^{2}}} = 0.2$, where $\phi$ is the azimuthal angle in radians.
This radius is reduced to $\Delta R = \max(0.05, 10\GeV / \pt$) for a lepton with $\pt > 50\GeV$.
Leptons are considered isolated if the scalar \pt sum within this radius is less than 10\% of the lepton \pt. Additionally, leptons are required to have a scalar \pt sum within a fixed radius of $\Delta R = 0.3$ less than $5\GeV$.  
Typical lepton selection efficiencies are approximately 85\% for electrons and 95\% for muons, depending on the \pt and $\eta$ of the lepton.

Events containing a second lepton passing a looser ``veto lepton'' selection, a \PGt passing a ``veto tau'' selection, or an isolated charged PF candidate are rejected. Hadronic \PGt decays are identified by a multi-variate analysis (MVA) isolation algorithm that selects both one-- and three--pronged topologies and allows for the presence of additional neutral pions~\cite{Khachatryan:2015dfa,CMS:2016gvn}. These vetoes are designed to provide additional rejection against events containing two leptons, or a lepton and a hadronic \PGt decay.

Hadronic jets are reconstructed from neutral and charged PF candidates associated with the primary vertex, using the anti-\kt clustering algorithm~\cite{Cacciari:2008gp,Cacciari:2011ma}.
Two collections of jets are produced, with different values of the distance parameter $R$.
Both collections of jets are corrected for contributions from event pileup and the effects of nonuniform detector response~\cite{Chatrchyan:2011ds}.

``Small-$R$'' jets are reconstructed with a distance parameter $R = 0.4$, and aim to reconstruct jets arising from a single parton.
Selected small-$R$ jets have $\pt > 30\GeV$, $\abs{\eta} < 2.4$, and are separated from isolated leptons by $\Delta R > 0.4$.
Small-$R$ jets that contain the decay of a \PQb-flavored hadron are identified as bottom quark jets (\PQb-tagged jets) using a deep neural network algorithm, \textsc{DeepCSV}. The discriminator working point is chosen so that the misidentification rate to tag light-flavor or gluon jets is approximately 1--2\%. This choice results in an efficiency to identify a bottom quark jet in the range 65--80\% for jets with \pt between 30 and 400\GeV, and an efficiency of 10--15\% for jets originating from a charm quark. The \PQb tagging efficiency in simulation is corrected using scale factors derived from comparisons of data with simulation in control samples~\cite{Sirunyan:2017ezt}.

\begin{table*}[bh]
\centering
  \setlength{\extrarowheight}{.7em}
\topcaption{
Summary of the requirements for the physics objects used in this analysis.}
\label{tab:objs}
    \begin{tabular}{ ll }
      \hline
      \multirow{2}{*}{Lepton} & $\ell = \PGm (\Pe)$ with $\pt^{\ell} > 25 (30)\GeV$, $\abs{\eta^{\ell}} <2.1~(1.44)$ \\
      & $\ptsum < 0.1 \: \pt^{\ell}$,  $\ptsum < 5\GeV$ \\
      [\cmsTabSkip]
      \multirow{2}{*}{Veto lepton} & $\PGm$ or $\Pe$ with $\pt^{\ell} > 5\GeV$, $\abs{\eta^{\ell}} <2.4$ \\
      & $\ptsum < 0.2 \: \pt^{\ell}$ \\
      \multirow{2}{*}{Veto track} & charged PF candidate, $\pt > 10\GeV$, $\abs{\eta} <2.4$ \\
      & $\ptsum < 0.1 \: \pt$,  $\ptsum < 6\GeV$ \\
      [\cmsTabSkip]

      \multirow{2}{*}{Veto \tauh} & hadronic \tauh with $\pt > 10\GeV$, $\abs{\eta} <2.4$ \\

      & \tauh MVA isolation \\
      [\cmsTabSkip]
      \multirow{2}{*}{Jets} & anti-\kt jets, $R = 0.4$, $\pt > 30\GeV$, $\abs{\eta} <2.4$ \\ 
      & anti-\kt jets, $R = 0.8$, $\pt > 250\GeV$, $\abs{\eta} <2.4$ \\

      \PQb tagging & \textsc{DeepCSV} algorithm (1\% misidentification rate) \\

      \PH tagging & mass-decorrelated \PH tagging discriminator \\
    
      \multirow{2}{*}{$\ptsum$ cone size} & $\ell$ relative isolation: $\Delta R = \min[ \max(0.05, 10\GeV / \pt^\ell), 0.2 ]$ \\
      & veto track, and $\ell$ absolute isolation: $\Delta R = 0.3$ \\
      [\cmsTabSkip]

      \hline
    \end{tabular}
\end{table*}

When the \pt of the Higgs boson is not too large compared to its mass, the \PQb jets resulting from its decay to bottom quarks are spatially separated.
As the Higgs boson \pt increases, the separation between the \PQb jets decreases. 
For the SUSY signal, this becomes important when the mass splitting between the neutralino \PSGczDt and the LSP is large.
To improve the sensitivity to large \PSGczDt masses, a second collection of ``large-$R$'' jets is formed with distance parameter $R = 0.8$.

Selected large-$R$ jets have $\pt >  250\GeV$, $\abs{\eta} < 2.4$, and are separated from isolated leptons by $\Delta R > 0.8$.
Large-$R$ jets containing a candidate $\PH\to\PQb\PAQb$ decay are identified as \PH-tagged jets using a dedicated deep neural network algorithm~\cite{Sirunyan:2020lcu}.
We use the mass-decorrelated version of the \textsc{DeepAK8} algorithm, which considers the properties of jet constituent particles and secondary vertices.
The imposed requirement on the neural network score corresponds to a misidentification rate of approximately 2.5\% for large-$R$ jets with a \pt of 500--700\GeV without an $\PH\to\PQb\PAQb$ decay in multijet events.
The efficiency to identify an \PH decay to bottom quarks is 60--80\% depending on the \pt of the large-$R$ jet.

The \ptvecmiss is modified to account for corrections to the energy scale of the reconstructed jets in the event.
Events with possible \ptvecmiss contributions from beam halo interactions or anomalous noise in the calorimeter are rejected using dedicated filters~\cite{Chatrchyan:2011tn}.
Additionally, during part of the 2018 data-taking period, two sectors of the endcap hadronic calorimeter experienced a power loss, affecting approximately 39\fbinv of data.
As the identification of both electrons and jets depends on correct energy fraction measurements, events from the affected data-taking periods containing an electron or a jet in the region $-2.4<\eta<-1.4$ and $-1.6<\phi<-0.8$ are rejected.
The total loss in signal efficiency considering all event filters is less than 1\%.

Data events are selected using a logical ``or'' of triggers that require either the presence of an isolated electron or muon; or large \ptmiss and \mht, where \mht is the magnitude of the negative vector \pt sum of all jets and leptons.
The combined trigger efficiency, measured with an independent data sample of events with a large scalar \pt sum of small-$R$ jets, is greater than 99\% for events with $\ptmiss>225\GeV$ and lepton $\pt>20\GeV$.
The trigger requirements are summarized in Table~\ref{tab:trigs}.

\begin{table*}[th]
\centering
  \setlength{\extrarowheight}{.7em}
\topcaption{
Summary of the triggers used to select the analysis data set. Events are selected using a logical ``or'' of the following triggers.
}
\label{tab:trigs}
    \begin{tabular}{ l }
      \hline
       $\ptmiss>120\GeV$ and $\mht>120\GeV$ (2016--2018) \\
       $\ptmiss>170\GeV$ (2016) \\
       Isolated $\PGm (\Pe)$ with $\pt^{\ell} > 24\: (25)\GeV$ (2016)\\
       Isolated $\PGm (\Pe)$ with $\pt^{\ell} > 24\: (35)\GeV$ (2017--2018)\\
      \hline
    \end{tabular}

\end{table*}

Table~\ref{tab:sels} defines the event preselection common to all signal regions, which requires exactly one isolated lepton, $\ptmiss > 125\GeV$, two or three small-$R$ jets, and no isolated tracks or veto tau candidates.

Exactly two of the small-$R$ jets must be \PQb-tagged.
The primary SM processes that contribute to the preselection region are \ttbar, single top quark (mostly in the $\PQt\PW$ channel), and \wjets production.

The SM processes with one \PW boson that decays to leptons, originating primarily from semileptonic \ttbar and \wjets, are suppressed by requiring the transverse mass, \mT, to be greater than $150\GeV$. \mT is defined as
\begin{equation}\label{eq:mt}
\mT = \sqrt{2 \pt^\ell\ptmiss (1 - \cos \Delta\phi)},
\end{equation}
where $\pt^\ell$ denotes the lepton $\pt$ and $\Delta\phi$ is the azimuthal separation between $\ptvec^\ell$ and \ptvecmiss.
After requiring a large \mT, the dominant remaining background comes from processes with two \PW bosons that decay to leptons (including \PGt leptons), primarily $\ttbar$ and \PQt{}\PW{}. To suppress these backgrounds, events with an additional veto lepton or a hadronic \PGt decay are rejected, as described above.

Additional background rejection is obtained using the cotransverse mass variable, \mct, which is defined as
\begin{equation}\label{eq:mct}
\MCT = \sqrt{2 \pt^{\PQb_1} \pt^{\PQb_2} (1 + \cos(\Delta\phi_{\PQb\PAQb}))},
\end{equation}
where $\pt^{b_1}$ and $\pt^{b_2}$ are the magnitudes of the transverse momenta of the two \PQb-tagged jets and $\Delta\phi_{\PQb\PAQb}$ is the azimuthal angle between the two \PQb-tagged jets~\cite{Tovey:2008ui}.
This variable has a kinematic endpoint close to $150\GeV$ for \ttbar events when both \PQb jets are correctly identified, while signal events tend to have higher values of \MCT.
Requiring $\MCT>200\GeV$ is effective at reducing the dilepton \ttbar and \PQt{}\PW{} backgrounds.

\begin{table*}[th]
\centering
   \setlength{\extrarowheight}{.7em}
\topcaption{Summary of the preselection requirements common to all signal regions.
The $\Nb$ is the multiplicity of {\PQb}-tagged jets and $\pt^{\mathrm{non\mbox{-}\PQb}}$ is the \pt of the non-{\PQb}-tagged jet.
}
\label{tab:sels}
\begin{tabular}{ ll}
\hline
      Lepton & Single $\Pe$ or $\PGm$ and no additional veto lepton, track or tau \\
      Small-$R$ jets & $2\leq \njets \leq 3$, $\Nb = 2$, $\pt^{\mathrm{non\mbox{-}\PQb}} < 300\GeV $\\
      \ptmiss & $>$125\GeV\\
      \mbb &  90--150\GeV \\
      \mT & $>$150\GeV \\
      \mct & $>$200\GeV\\
      \hline

  \end{tabular}
\end{table*}

\begin{figure}[htb]
\centering
\includegraphics[width=0.32 \textwidth]{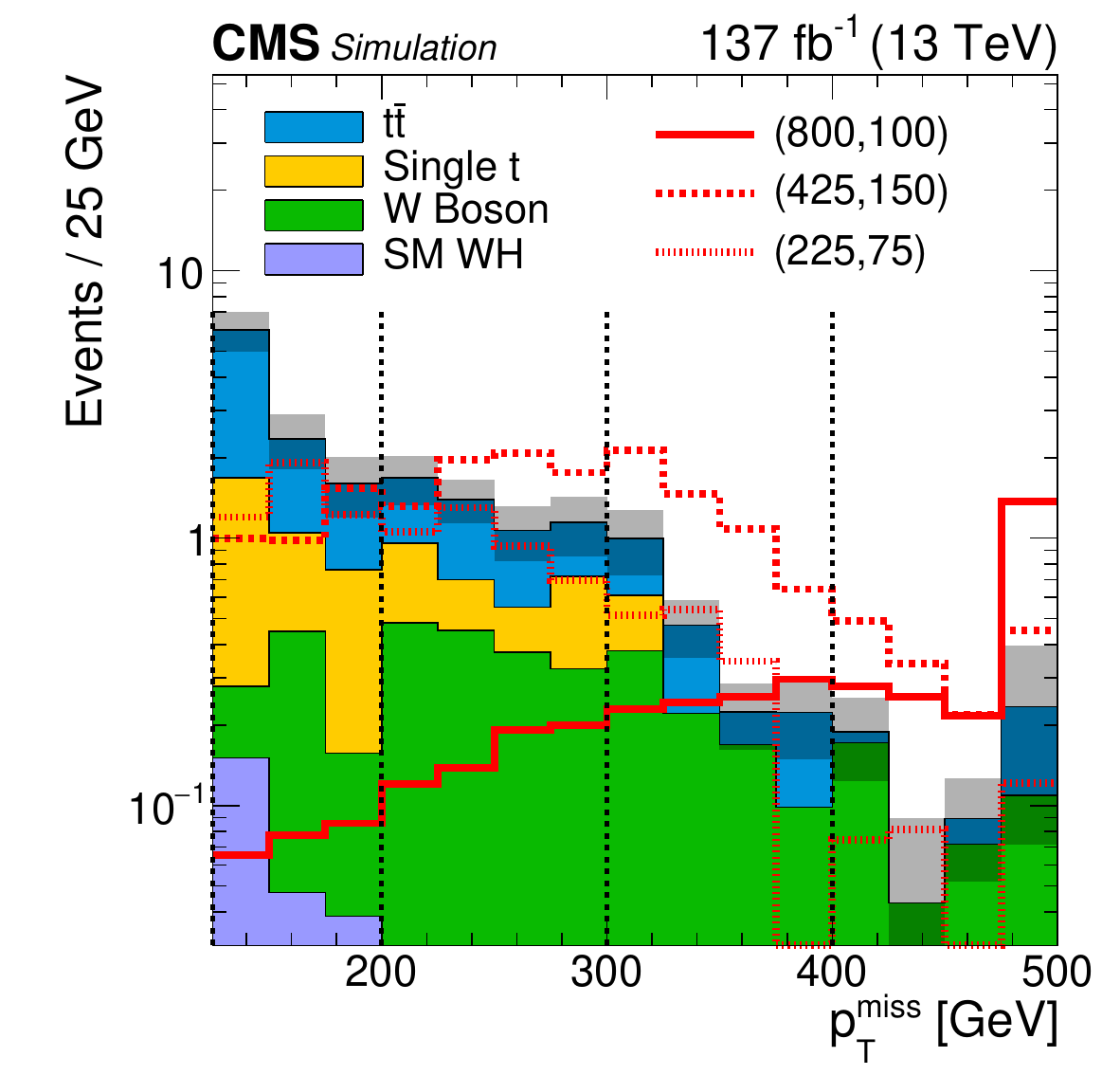}
\includegraphics[width=0.32 \textwidth]{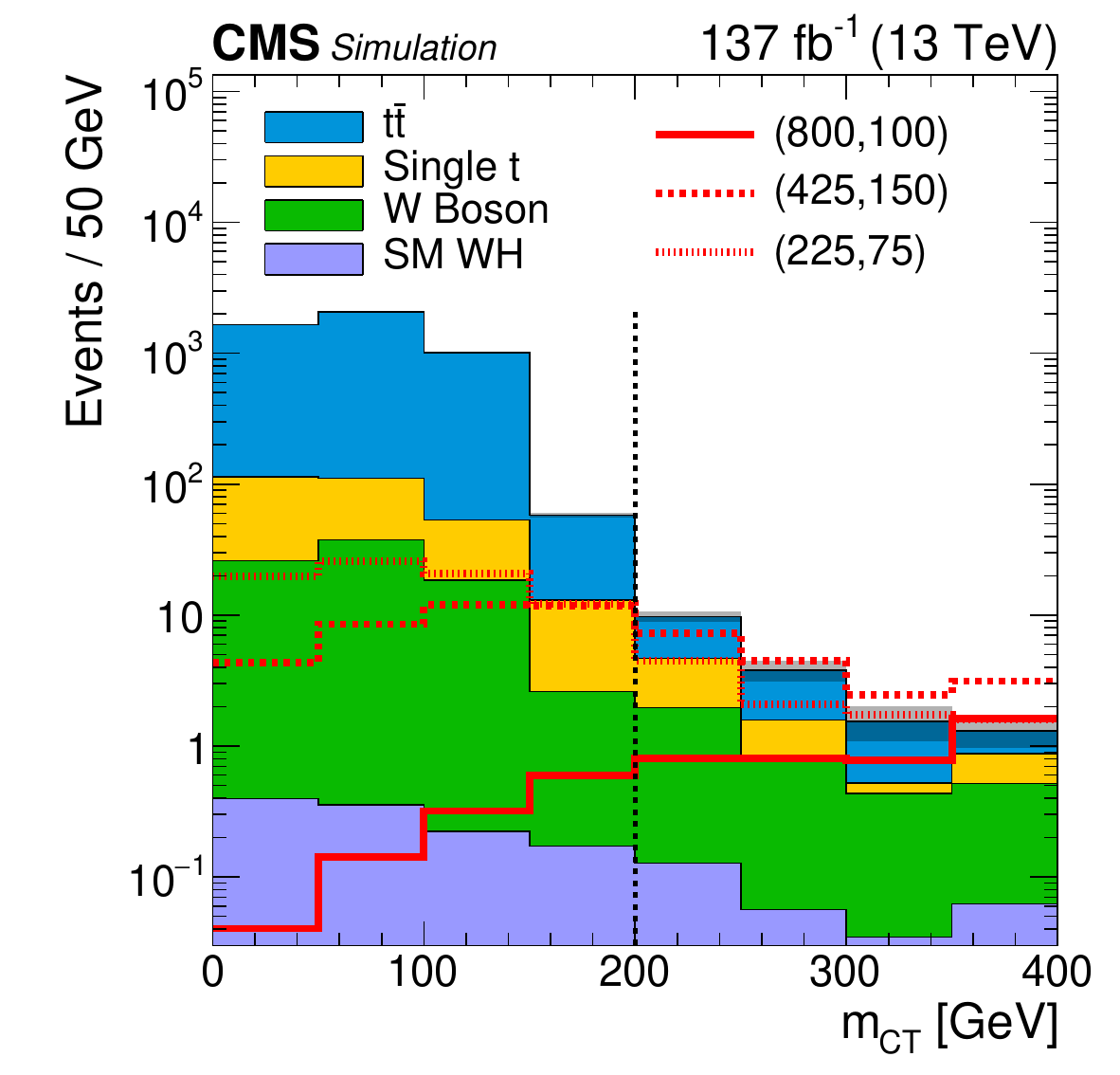}
\includegraphics[width=0.32 \textwidth]{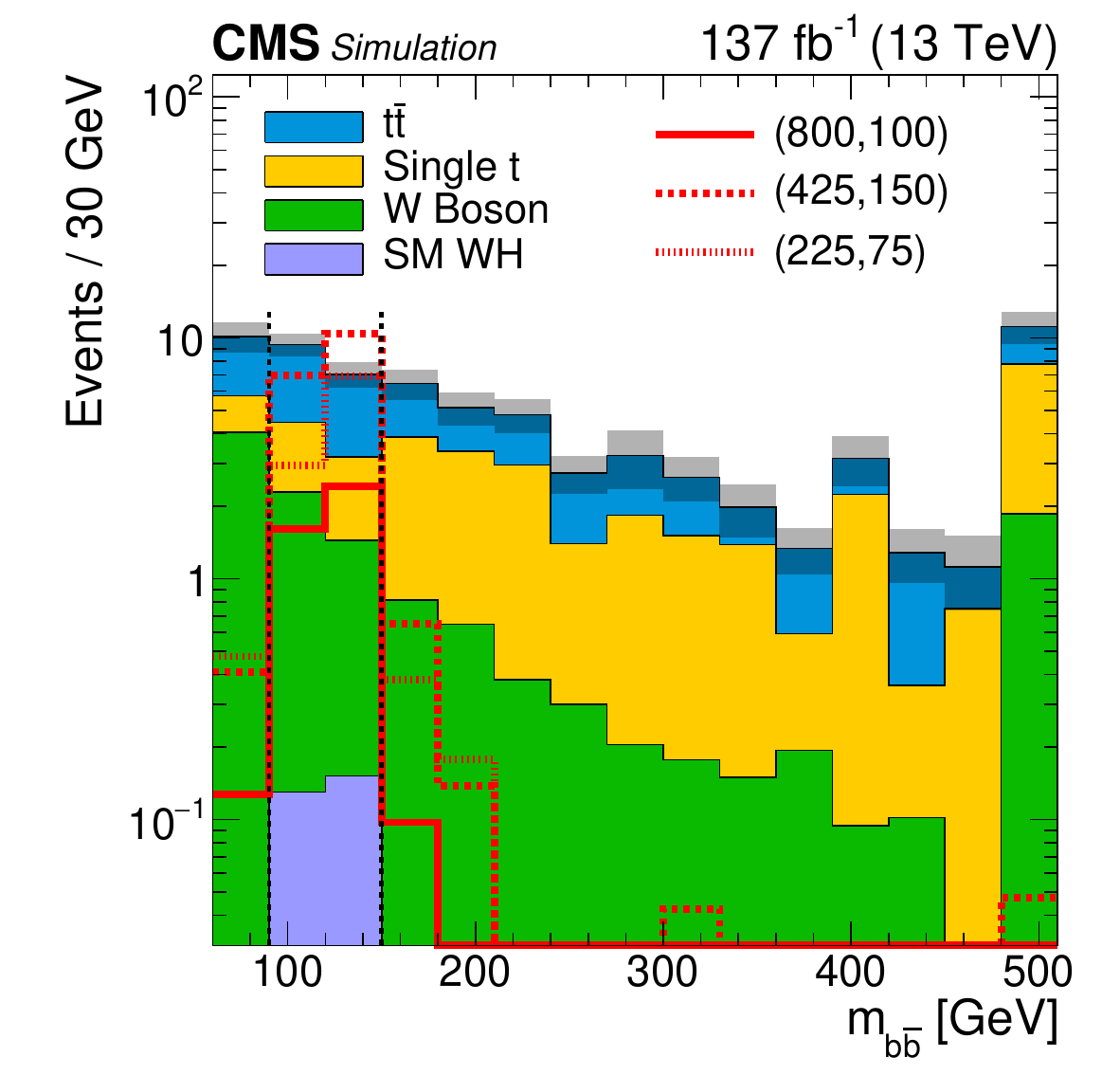}\\
\includegraphics[width=0.32 \textwidth]{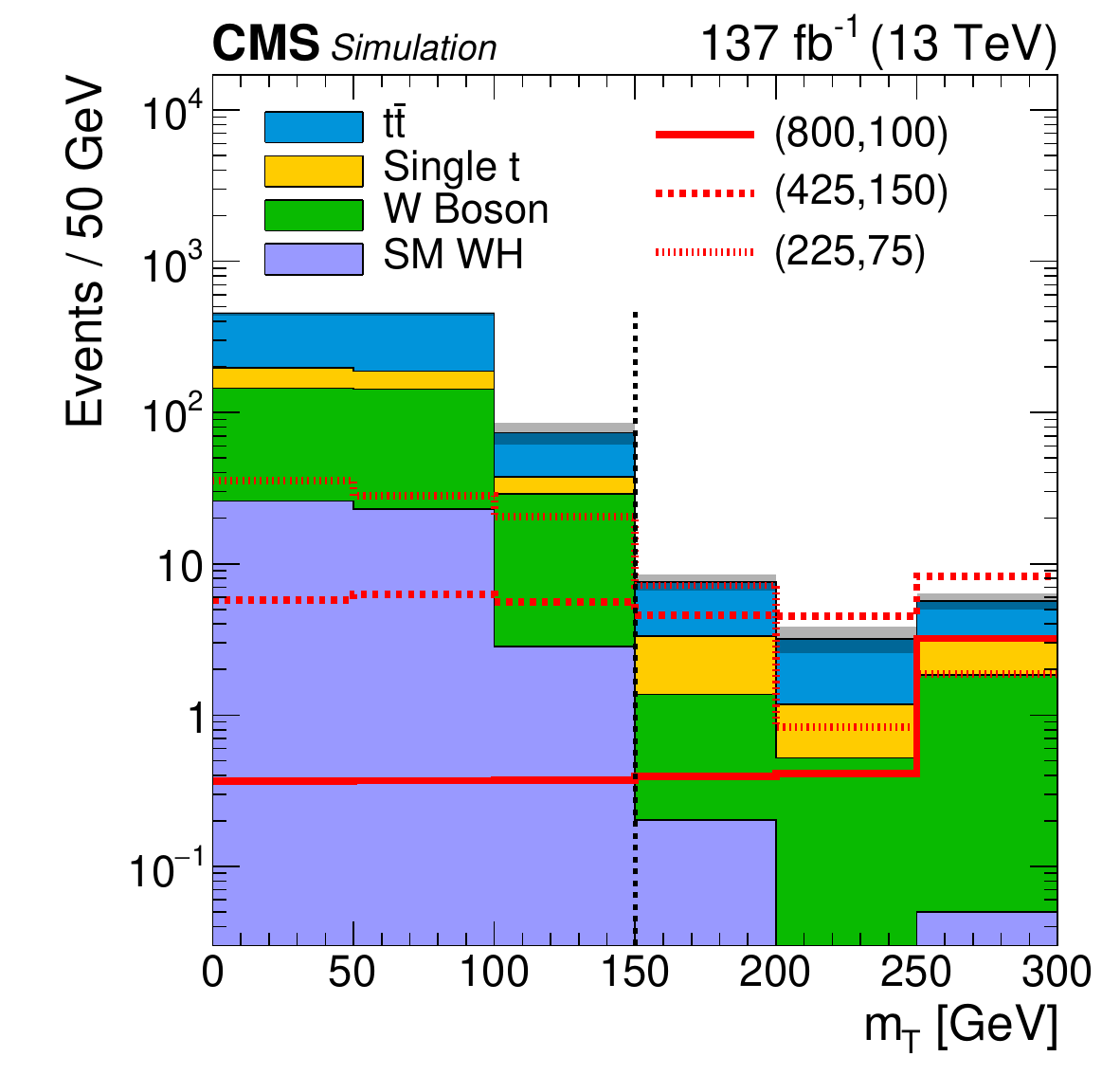}
\includegraphics[width=0.32 \textwidth]{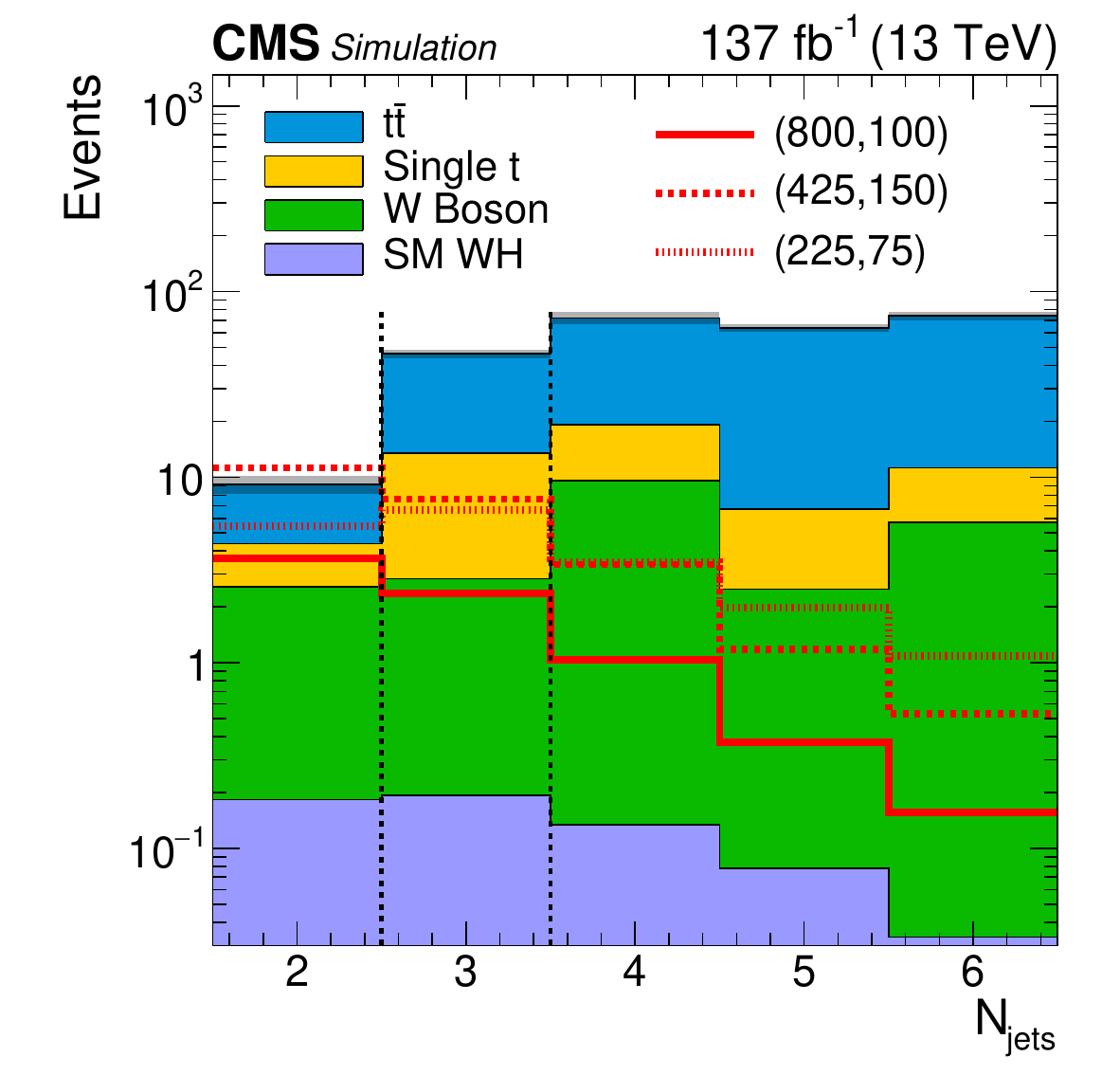}
\includegraphics[width=0.32 \textwidth]{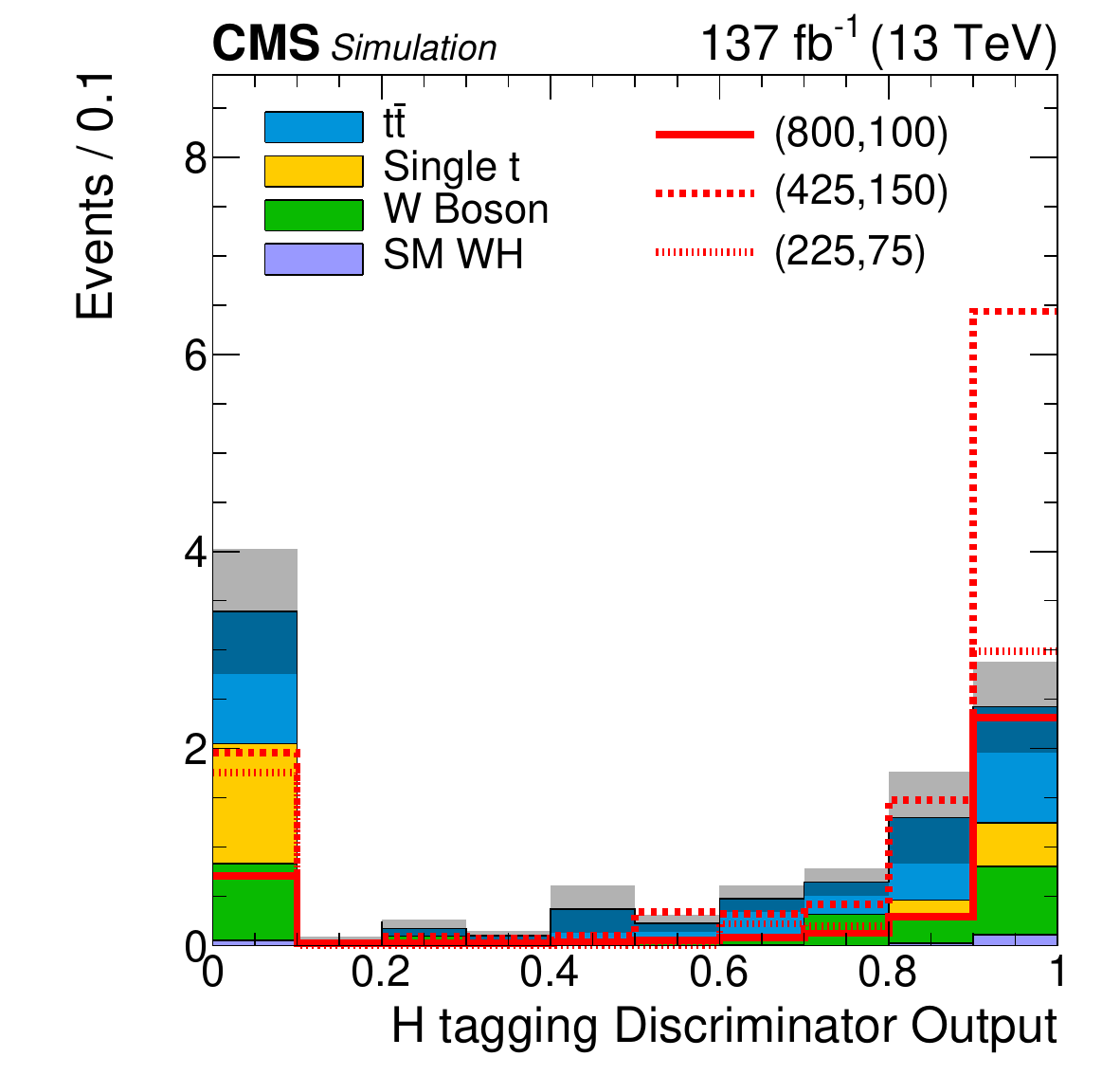}
\caption{
Distributions of \ptmiss, \mct, \mbb, \mT, \njets, and the $\PH\to\PQb\PAQb$ large-$R$ jet discriminator in simulated background and signal samples.
Three benchmark signal points corresponding to masses in GeV ($m_{\PSGczDt/\PSGcpmDo}$, $m_{\PSGczDo}$) of (800, 100), (425, 150) and (225, 75) are shown as solid, dashed, and short-dashed lines, respectively.
Events are taken from the 2-jet signal regions with $\ptmiss > 125\GeV$, with all of the requirements specified in Table~\ref{tab:sels} except for the one on the plotted variable.
The shaded areas correspond to the statistical uncertainty of the simulated backgrounds. The dashed vertical lines indicate the thresholds used to define the signal regions.
These indicators are not shown on the \PH tagging discriminator score distribution because the required values vary between 0.83 and 0.90, depending on the data-taking year.
}
\label{fig:presel}
\end{figure}

Events entering the signal regions must pass the preselection and satisfy the \mT and \MCT requirements above.
We also require that the invariant mass of the pair of \PQb-tagged jets, \mbb, be between 90 and 150\GeV, consistent with the mass of an SM Higgs boson.
In events with 3 small-$R$ jets, the non-\PQb-tagged jet must have $\pt < 300\GeV$. This requirement rejects some \ttbar events that survive the \mct and \ptmiss selections.
These requirements define the baseline signal selection.
Figure~\ref{fig:presel} shows the distributions of \ptmiss, \mct, \mbb, \mT, the number of small-$R$ jets (\njets), and the discriminator output of the \PH tagging algorithm in simulated signal and background samples. All preselection requirements specified in Table~\ref{tab:sels} are applied except the one on the plotted variable, illustrating the discrimination power of each variable.

Events passing the baseline signal selection are further categorized into signal regions according to \njets, the number of \PH-tagged large-$R$ jets \nhiggs, and the value of \ptmiss.
The twelve non-overlapping signal regions are defined in Table~\ref{tab:bins}.

\begin{table*}[th]
\centering
   \setlength{\extrarowheight}{.7em}
\topcaption{Definition of the 12 non-overlapping signal regions categorized in \nH, \njets, and \ptmiss, where $\nH$ is the number of large-$R$ jets tagged as $\PH\to\PQb\PAQb$.}
\label{tab:bins}

    \begin{tabular}{ lll }
      \nH &\njets &\ptmiss [GeV] \\
      \hline
      0 & 2, 3 & [125, 200), [200, 300), [300, 400), [400, $\infty$)\\
      1 & 2, 3 & [125, 300), [300, $\infty$)\\

      \hline

    \end{tabular}
\end{table*}

\section{Background estimation}
\label{Sec:BkgEst}

There are two dominant background categories relevant for this search: top quark production and \PW boson production. 
The contributions of these backgrounds to the yields in the signal regions are estimated using observed yields in control regions (CRs) and transfer factors obtained from simulated samples. The transfer factors are validated in non-overlapping regions adjacent to the signal regions.
The top quark backgrounds include \ttbar pair production, single top quark production (\PQt{}\PW), and a small contribution from $\ttbar\PW$ and $\ttbar\PZ$ production. These backgrounds dominate in the lower-\ptmiss search regions and are estimated from CRs in data using the method described in Section~\ref{ssec:topbg}. In the high-\ptmiss regions, \PW boson production becomes the dominant background. The method described in Section~\ref{sec:wjetsbg} estimates the background arising from \wjets, \PW{}\PW, and \PW{}\PZ production using CRs in data. The remaining background arises from standard model \PW{}\PH production. This process contributes less than 5\% of the total background in any of the search regions, and its yield is estimated from simulation. A 25\% uncertainty in the cross section of this process is assigned, based on the uncertainty in the \PW{}\PH cross section measurement~\cite{Sirunyan:2018kst}.

\subsection{Top quark background}
\label{ssec:topbg}

Events containing top quarks constitute the dominant background, particularly in bins with $\njets = 3$ or low \ptmiss. These events contain \PQb jets and isolated leptons from \PW bosons, so they lead to similar final states as the signal. Owing to the high \mT requirement, the majority of the top quark background stems from events with two leptonically decaying \PW bosons. In this case, one of the leptons either is not reconstructed, fails the identification requirements, is not isolated, or is outside of kinematic acceptance.

The \ttbar background is further suppressed by the \mct requirement, which has an endpoint at approximately $150\GeV$ for \ttbar events in the case when both daughter \PQb jets are reconstructed and identified. The \mct value for \ttbar events can exceed the cutoff for three reasons: (i) if there are mistagged light-flavor jets or extra \PQb jets, (ii) if a \PQb jet is reconstructed with excess \pt because it overlaps with other objects, or (iii) because of excess \PQb jet \pt arising due to the finite jet energy resolution.

A control sample enriched in top quark events is obtained by inverting the \mct requirement. For each signal region (SR), we form a corresponding control region spanning a range of \mct from $100$ to $200\GeV$.  
These CRs are used to normalize the top quark background to data in a single-lepton, high-\mT region in each bin of \ptmiss, \nhiggs, and \njets. 
In each CR, a transfer factor from MC simulation (\rmct) is used to extrapolate the yield for the corresponding high-\mct signal regions. The top quark background estimate is then given by
\begin{equation}
\label{eq:rmctMethod}
  N_{\mathrm{SR}}^{\text{top}}(\ptmiss,\njets,\nh) = \rmct(\ptmiss,\njets,\nh) N_{\mathrm{CR}}^{\text{obs.}}(\ptmiss,\njets,\nh), 
\end{equation}
where the $N_{\mathrm{SR}}^{\text{top}}$ is the number of expected events in the SR, $N_{\mathrm{CR}}^{\text{obs.}}$ is the number of observed events in the CR, and \rmct are defined as
\begin{equation}
\label{eq:rtop}
  \rmct(\ptmiss,\njets,\nh) =  \frac{N_{\mathrm{SR}}^{\text{top}~\mathrm{MC}}(\ptmiss,\njets,\nh)} {N_{\mathrm{CR}}^{\mathrm{SM}~\mathrm{MC}}(\ptmiss,\njets,\nh)}. 
\end{equation}
The $N_{\mathrm{SR}}^{\text{top}~\mathrm{MC}}$ and $N_{\mathrm{CR}}^{\mathrm{SM}~\mathrm{MC}}$ are the expected top quark and total SM yields in the signal and control regions, respectively, according to simulation.

The contamination from other processes (primarily \PW boson production) in the low-\mct CRs is as low as 2\% in the lower-\ptmiss regions, growing to 25\% in the highest \ptmiss control region. This contamination is included in the denominator of \rmct as shown in Eq.~(\ref{eq:rtop}).
Additionally, to increase the expected yields in the CRs, two modifications to the CR definitions are made. First, for the CRs with an \PH-tagged large-$R$ jet, the \mct lower bound is removed (for a total range of $0$--$200\GeV$). Second, for CRs with $\ptmiss > 300\GeV$, the \mbb window is expanded to 90--300\GeV.

The data yields, transfer factors, and the resulting top quark background predictions are summarized in Table~\ref{tab:mct_yields}. These predictions, combined with the other background estimates, are compared with the observed yields in Section~\ref{sec:results}.

\begin{table}[htbp!]{}
  \centering
  \topcaption{The values of the \rmct transfer factors, the observed yields in the low-\mct CRs, and the resulting top quark background prediction in each bin of \ptmiss, \njets, and \nh. The uncertainty shown for \rmct is only of statistical origin. For the top quark prediction both the statistical and systematic uncertainties are shown (discussed in the text.)}
  \label{tab:mct_yields}

  \begin{tabular}{cccccc}

   \njets & \nh & \ptmiss [\GeVns{}]   & \rmct & $N_{\mathrm{CR}}^{\text{obs.}}$  &  $N_{\mathrm{SR}}^{\text{top}}$ \\ \hline
   \multirow{6}{*}{2} & \multirow{4}{*}{0}
   & 125--200  & $0.006 \pm 0.001$ &   $978$ & $6.3 \pm 0.9 \pm 0.9$    \\
   &&200--300  & $0.015 \pm 0.003$ &  $161$ & $2.4 \pm 0.5 \pm 0.4$     \\
   &&300--400  & $0.05 \pm 0.02$&  $6$   & $0.3 \pm 0.1 \pm 0.1$    \\
   &&$>$400    & $0.02 \pm 0.02$&  $1$   & $0.02 \pm 0.02 \pm 0.01$     \\ [1\cmsTabSkip]
   & \multirow{2}{*}{1}
   & 125--300  & $0.26 \pm 0.06$&   $6$   & $1.6 \pm 0.8 \pm 0.4$    \\
   &&$>$300    & $0.03 \pm 0.01$ &  $11$   & $0.4 \pm 0.2 \pm 0.3$    \\ [2\cmsTabSkip]

   \multirow{6}{*}{3}  & \multirow{4}{*}{0}
   & 125--200   & $0.020  \pm 0.002$ &$851$ & $17.5 \pm 1.6 \pm 2.6$   \\
   && 200--300  & $0.05 \pm 0.01$ &$151$ &  $7.1 \pm 1.1 \pm 1.3$   \\
   && 300--400  & $0.04 \pm 0.01$&$19$  &  $0.8 \pm 0.3 \pm 0.3$    \\
   &&$>$400     & $0.2 \pm 0.2$ &$1$   &  $0.2 \pm 0.2 \pm 0.1$     \\ [1\cmsTabSkip]
   & \multirow{2}{*}{1}
   & 125--300   & $0.28  \pm 0.05$ &$18$  &  $5.0 \pm 1.4 \pm 1.4$    \\
   &&$>$300     & $0.12  \pm 0.03$ &$14$  &  $1.7 \pm 0.7 \pm 1.4$    \\
   \end{tabular}

\end{table}

To assess the modeling of the top quark background, we conduct a validation test in a sideband requiring $\mbb > 150\GeV$ and the same \mct and \mT requirements as the SR. 
The relative contributions from SM processes are similar in the sideband and the signal regions.
The modeling of the top quark background in this region is also affected by the same sources of uncertainty, including the
imperfect knowledge of the object efficiencies, jet energy scale and resolution, and the distribution of additional pileup interactions. An analogous background prediction is performed in this region, and the level of agreement observed is used to derive a systematic uncertainty in the \rmct factors.

The yields in the $\mbb > 150\GeV$ validation regions (VRs) are estimated using CRs defined with the same \mT and \mct requirements as the CRs for the SR predictions: $\mT > 150\GeV$, and $\mct > 100$ $(0) \GeV$ for $\nh = 0$ $(1)$.  Two modifications are introduced to improve the statistical precision of the test: first, the $\njets=2$ and $\njets=3$ bins are combined; and second, all regions with $\ptmiss > 300$\GeV and $\ptmiss > 400$\GeV are combined. Additionally, to avoid overlap with the low-\mct control regions used to estimate the top quark background in the SR, the low-\mct regions used for the VR predictions in bins with $\ptmiss > 300 \GeV$ are restricted to $\mbb > 300 \GeV$.
 
\begin{figure}[th]
\centering
\includegraphics[width = .99\textwidth]{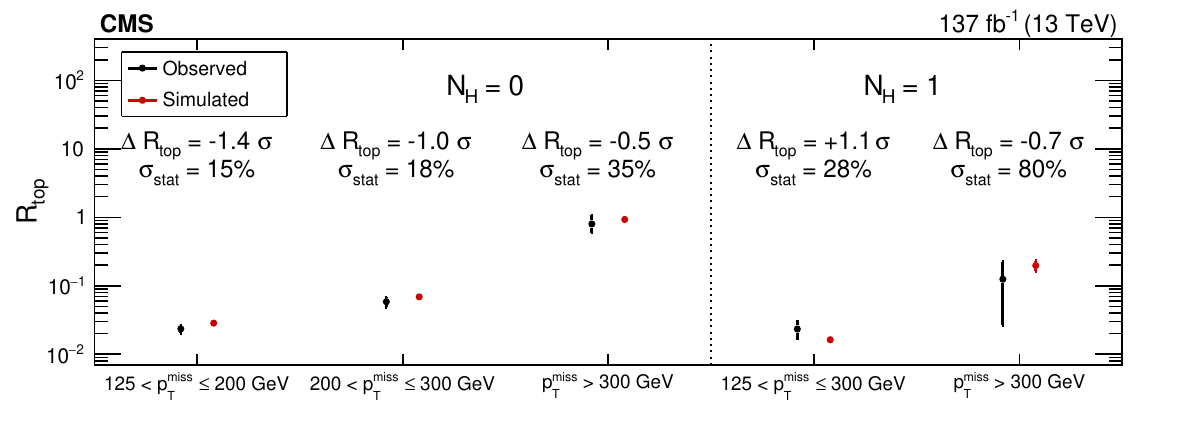}

\caption{Observed and simulated \rmct values in the $\mbb > 150\GeV$ validation regions. The differences between observed and simulated \rmct values, divided by the total statistical uncertainties, are also listed in the figure as $\Delta \rmct$. The statistical precision of each difference, $\sigma_{\mathrm{stat}}$, is taken as the systematic uncertainty on \rmct for the corresponding bin in the signal region. 
}
\label{fig:ratio_mbbtest_vs_met}
\end{figure}

A comparison of the \rmct factors obtained from data and simulation in the VRs is shown in Fig.~\ref{fig:ratio_mbbtest_vs_met}. Good agreement is observed, and we assign the statistical uncertainties in the differences of the observed and simulated values as the systematic uncertainties in the corresponding \rmct factors. These uncertainties reflect the degree to which we can evaluate the modeling of \rmct factors in data. This validation approach has the advantage of probing both the known sources of uncertainty as well as any unknown sources that could affect the \mct extrapolation. The uncertainties derived from this test, together with those associated with the finite yields in the low-\mct CRs and the MC statistical precision form the complete set of uncertainties assigned to the top quark background prediction.

Additional cross-checks of the top quark background estimate are performed in a dilepton validation region and in a region with exactly one \PQb jet.
These studies are performed in all 12 bins of \ptmiss, \njets, and \nH, and 
the results agree with those obtained from the studies performed in the \mbb sideband.
A second, independent estimate of the top quark background is performed following the ``lost-lepton'' method described in Ref.~\cite{Sirunyan_2020_032}.
In this method, the contribution from top quark processes in each signal region is normalized using a corresponding control region requiring two leptons and all other signal region selections.
The estimates obtained from the two methods are consistent.
These additional cross-checks are not used quantitatively to determine uncertainties, but they build confidence in the modeling of the \rmct factors.

\subsection{\texorpdfstring{\PW}{W} boson background}
\label{sec:wjetsbg}

Events arising from \PW boson production, mainly \wjets, \PW{}\PW{}, and \PW{}\PZ{}, are the second largest background in this search and are the dominant SM contribution in bins with high \ptmiss.
Events from \wjets production satisfy the baseline selection when they contain true \PQb jets originating from $\Pg\to\bbbar$ (associated \PW production with heavy-flavor jets, \whfjets) or when light-flavor jets are misidentified as \PQb jets (associated \PW production with light flavor jets, \wlfjets).
Because of the low misidentification rate of light-flavor jets, more than 75\% of the selected \wjets events contain at least one genuine \PQb jet.
The \wjets background is reduced by the $\mT>150\GeV$ requirement.
In absence of large mismeasurements of the \ptmiss, the \PW boson must be produced off-shell in order to satisfy this threshold.

The \PW boson background is normalized in a data control sample obtained by requiring the number of \PQb-tagged jets (\Nb) to be less or equal to 1 and the same \mT, \mct, and \mbb requirements as the signal regions.
The $\Nb=0$ region of this sample is used to normalize the \PW boson background while the $\Nb=1$ region is used to constrain the contamination from top quark events.
The two jets with the highest \PQb tagging discriminator values are used to calculate \mbb and \mct.
The control sample is binned in \njets and \ptmiss following the definition of the signal regions
and has a high purity of \PW boson events for $\Nb=0$.
 
The contribution from processes involving top quarks, mostly single or pair production of top quarks, is up to 20\% in some $\Nb=0$ CRs.
The contamination is estimated by fitting the \Nb distribution in each CR using templates of \wjets and top quark events obtained from simulation.
The templates are extracted from simulated \PW boson and top quark samples, respectively.
The number of \PW boson events in each CR, $N^{\PW}_{\CR}$, is obtained by subtracting from the observed yield, $N_{\CR}^{\text{obs.}}$, the contribution of top quark events $N^{\text{top}}_{\CR}$.
For the yield $N^{\text{top}}_{\CR}$, a correction factor obtained from the fit, which is typically close to 1.1, is taken into account.

We define a transfer factor \rw to extrapolate from each $\Nb=0$ CR to the corresponding $\Nb=2$ signal region.
Simulated samples of \PW boson processes are used to calculate \rw.
Since there are very few events with an \PH-tagged large-$R$ jet in the control samples, it is not feasible to form dedicated CRs with $\nhiggs=1$.
Instead, the control samples are inclusive in \nhiggs, and the extrapolation into $\nhiggs=0$ and $\nhiggs=1$ is handled by the \rw factors.
The predicted yield of the \PW boson background in each of the signal regions, $N^{\PW}_{\SR}$, is therefore given by
\begin{equation}
  N^{\PW}_{\SR}(\ptmiss, \njets, \nhiggs) = N^{\PW}_{\CR}(\ptmiss, \njets) \rw (\ptmiss, \njets, \nhiggs)
\end{equation}
with
\begin{equation}
  N^{\PW}_{\CR}(\ptmiss, \njets) =  N_{\CR}^{\text{obs.}} (\ptmiss, \njets) -  N^{\text{top}}_{\CR}(\ptmiss, \njets) ,
\end{equation}
and \rw is defined as
\begin{equation}
  \rw (\ptmiss, \njets, \nhiggs) = \frac{N^{\PW~\textrm{MC}}_{\SR} (\ptmiss, \njets, \nhiggs) }{N^{\PW~\textrm{MC}}_{\CR} (\ptmiss, \njets) } .
\end{equation}
The resulting predictions are shown in Table~\ref{tab:R_W}. Section~\ref{sec:results} shows a comparison with the observed yields after combining with the other background estimates.

\begin{table}[htbp]
  \centering
  \topcaption{
    The observed ($N_{\CR}^{\text{obs.}}$) and top quark background yield ($N_{\CR}^{\PW}$) in the \CR, together with the values of \rw for the extrapolation of the \PW boson background from the \CR to the \SR, and the final \PW boson prediction, $N^{\PW}_{\SR}$.
    The uncertainties in \rw include the statistical uncertainty only. 
    The \PW boson prediction shows both the statistical and systematic uncertainties.}
  \label{tab:R_W}
  \begin{tabular}{ccccccccc}

   \njets             & \ptmiss [GeV]  & $N_{\CR}^{\text{obs.}}$ & $N^{\text{top}}_{\CR}$ & $N_{\CR}^{\PW}$ & & \nhiggs             & $\rw \times 10^3$   & $N^{\PW}_{\SR}$ \\ \hline
   \multirow{6}{*}{2} & 125--200 & 449                       & $65\pm7$             & $384\pm23$      & & \multirow{4}{*}{0}  & $1.3 \pm 0.6$       & $0.5 \pm 0.2 \pm 0.1$ \\
                      & 200--300 & 314                       & $34\pm45$            & $280\pm19$      & &                     & $3.6 \pm 0.7$       & $1.0 \pm 0.2 \pm 0.2$ \\
                      & 300--400 & 191                       & $10\pm1$             & $181\pm14$      & &                     & $3.7 \pm 0.7$       & $0.7 \pm 0.1 \pm 0.1$ \\
                      & $>$400   & 110                       & $2.5\pm0.7$          & $108\pm11$      & &                     & $2.8 \pm 0.8$       & $0.3 \pm 0.1 \pm 0.1$ \\ [1\cmsTabSkip]
                      & 125--300 &                           &                      &                 & & \multirow{2}{*}{1}  & $1.1 \pm 0.2$       & $0.7 \pm 0.2 \pm 0.1$ \\
                      & $>$300   &                           &                      &                 & &                     & $1.7 \pm 0.7$       & $0.5 \pm 0.2 \pm 0.2$ \\ [2\cmsTabSkip]
   \multirow{6}{*}{3} & 125--200 & 329                       & $67\pm5$             & $262\pm19$      & & \multirow{4}{*}{0}  & $0.9 \pm 0.6$       & $0.2 \pm 0.2 \pm 0.1$ \\
                      & 200--300 & 152                       & $32\pm5$             & $120\pm14$      & &                     & $5.9 \pm 1.5$       & $0.7 \pm 0.2 \pm 0.1$ \\
                      & 300--400 & 81                        & $7\pm1$              & $74\pm10$       & &                     & $9.4 \pm 2.6$       & $0.7 \pm 0.2 \pm 0.2$ \\
                      & $>$400   & 44                        & $3.7\pm1.7$          & $40\pm7$        & &                     & $6.5 \pm 1.9$       & $0.3 \pm 0.1 \pm 0.1$ \\ [1\cmsTabSkip]
                      & 125--300 &                           &                      &                 & & \multirow{2}{*}{1}  & $2.0 \pm 0.5$       & $0.8 \pm 0.2 \pm 0.2$ \\
                      & $>$300   &                           &                      &                 & &                     & $2.9 \pm 1.7$       & $0.3 \pm 0.2 \pm 0.1$ \\
   \end{tabular}
\end{table}

To assess the modeling of heavy-flavor jets in the simulated \whfjets samples, we perform a similar extrapolation in \Nb in a Drell--Yan (DY) validation sample assuming $\PZ \to \ell \ell$.
The large contribution from \ttbar in the $\Nb=2$ region is suppressed by requiring two opposite-charge, same-flavor leptons with an invariant mass compatible with a \PZ boson, $\abs{m(\ell\ell)-m_{\PZ}}<5\GeV$.
In the validation sample, the predicted and observed \dyhfjets yields agree within 20\%.
Based on this test, we vary the fraction of \wjets events with at least one generated \PQb jet by 20\% and assign the resulting variation of \rw as a systematic uncertainty.

\begin{figure}[t]
\centering
\includegraphics[width=0.45 \textwidth]{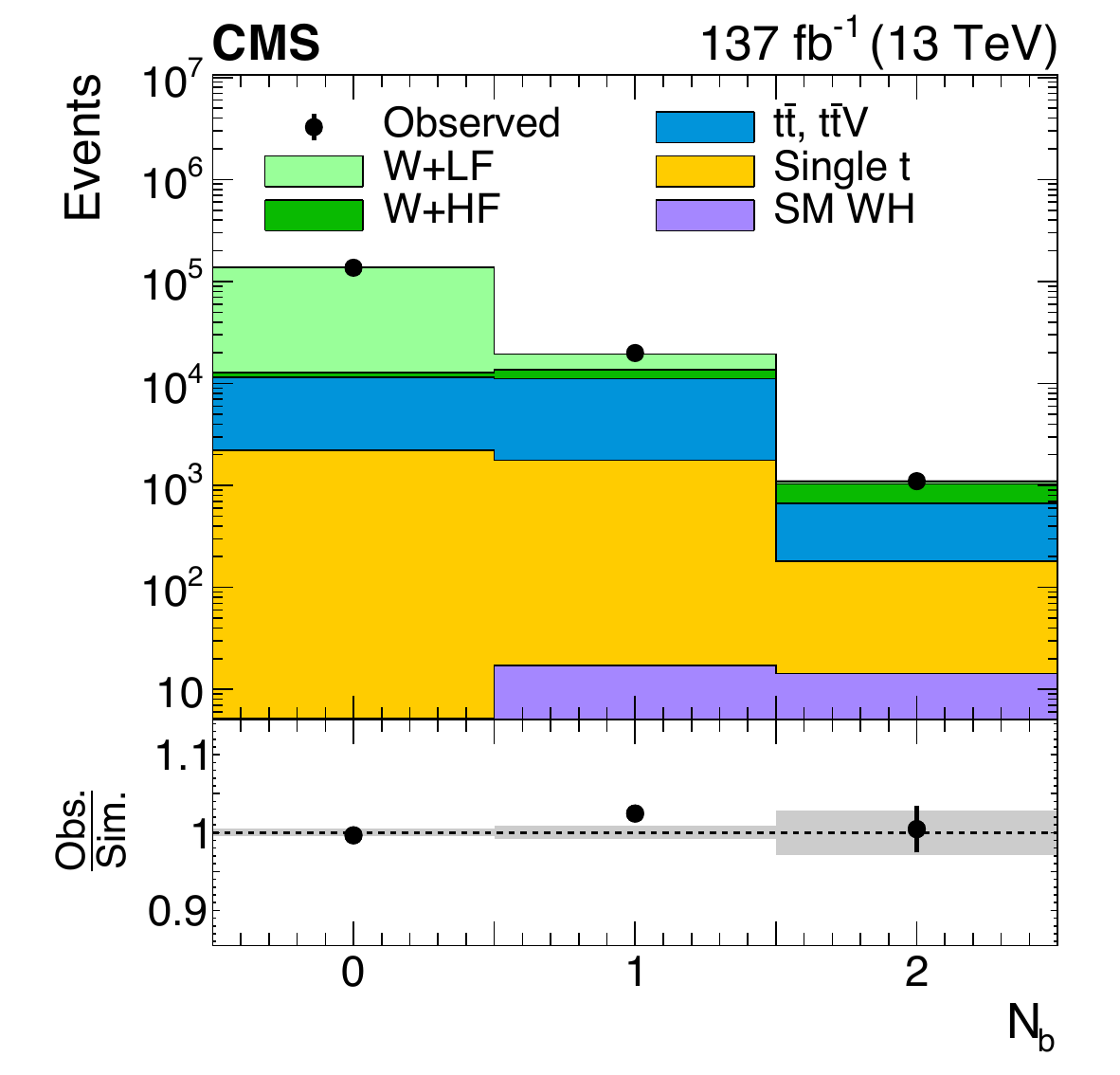}
\caption{
Distribution of \Nb in the low-\mT control sample. The \ttjets contribution is suppressed by requiring $\mct > 200\GeV$. The shaded area reflects the statistical uncertainty in the simulation.
}
\label{fig:wjets_btag}
\end{figure}

We also study the distribution of \Nb in a low-\mT control sample, obtained by selecting events with $\ptmiss>125\GeV$, $50 < \mT < 150\GeV$, $\njets=2$, and without a requirement on $\mbb$.
The top quark contribution in this region is largely suppressed by the $\mct>200\GeV$ requirement, yielding a sample with a \whfjets purity of approximately 40\% for $\Nb=2$.
Good agreement between data and simulation is observed in this region, as shown in Fig.~\ref{fig:wjets_btag}.

Additional contributions to the uncertainty in the factor \rw are evaluated.
The difference of the \whfjets fraction with respect to the one derived from the \dyhfjets validation test results in a systematic uncertainty of up to 16\% in \rw.
Based on the latest measurements~\cite{Sirunyan:2019bez,Sirunyan:2020jtq,Sirunyan:2020pub} and considering the delicate phase space requiring significant \ptmiss and $\Nb=2$, the diboson production cross section is varied by 25\%, yielding a maximum systematic uncertainty of 12\%.
The uncertainties from the measurement of the \PQb tagging efficiency scale factors are propagated to the simulated \wjets and diboson events resulting in an uncertainty of up to 10\% in \rw.
The simulated samples are reweighted according to the distribution of the true number of interactions per bunch crossing.
The uncertainty in the total inelastic \pp cross section results in uncertainties of 2--6\% in \rw.
The uncertainty arising from the jet energy calibration~\cite{CMS-DP-2020-019} is assessed by shifting jet momenta in simulated samples up and down, and propagating the resulting changes to \rw.
Typical values for the systematic uncertainty from the jet energy scale range from 2--10\%, reaching up to 20\% for events with a boosted Higgs boson candidate.

The mistag rate of the \PH tagging algorithm for large-$R$ jets that do not contain a true \PH is measured in 
a control sample obtained by requiring low-\mT, $\Nb = 2$, and at least one large-$R$ jet.
Scale factors are measured and applied to simulation to correct for differences in the observed mistag rates.
The uncertainty in the scale factors is dominated by the limited statistical precision of the control sample and results in a systematic uncertainty up to 14\% in \rw.

The renormalization ($\mu_{\mathrm{R}}$) and factorization ($\mu_{\mathrm{F}}$) scales are varied up and down by a factor of 2, omitting the combination of variations in opposite directions.
The envelope of the variations reaches values up to 15\% and is assigned as systematic uncertainty.
The uncertainties resulting from variations of the PDF and the strong coupling \alpS are less than 2\%.
The systematic uncertainties in \rw are summarized in Table~\ref{tab:WSys}.

\begin{table}[th]
    \centering
    \topcaption{Systematic uncertainties on \rw.}
    \begin{tabular}{lc}
        Source                                   & Typical values \\ \hline
        \whfjets fraction                        & 7--16\%      \\
        Diboson cross section                    & 1--12\%      \\
        \PQb tagging efficiency                  & 3--10\%      \\
        \PH mistag rate                          & 3--14\%      \\
        Jet energy scale                         & 2--20\%      \\
        Pileup                                   & 1--6\%       \\
        PDF                                      & $<$2\%       \\
        \alpS                             & $<$2\%       \\
        $\mu_{\mathrm{R}}$ and $\mu_{\mathrm{F}}$ & 3--15\%     \\
    \end{tabular}
    \label{tab:WSys}
\end{table}

\section{Results and interpretation}
\label{sec:results}

\begin{figure*}[htb]
\centering
\includegraphics[width=0.65 \textwidth]{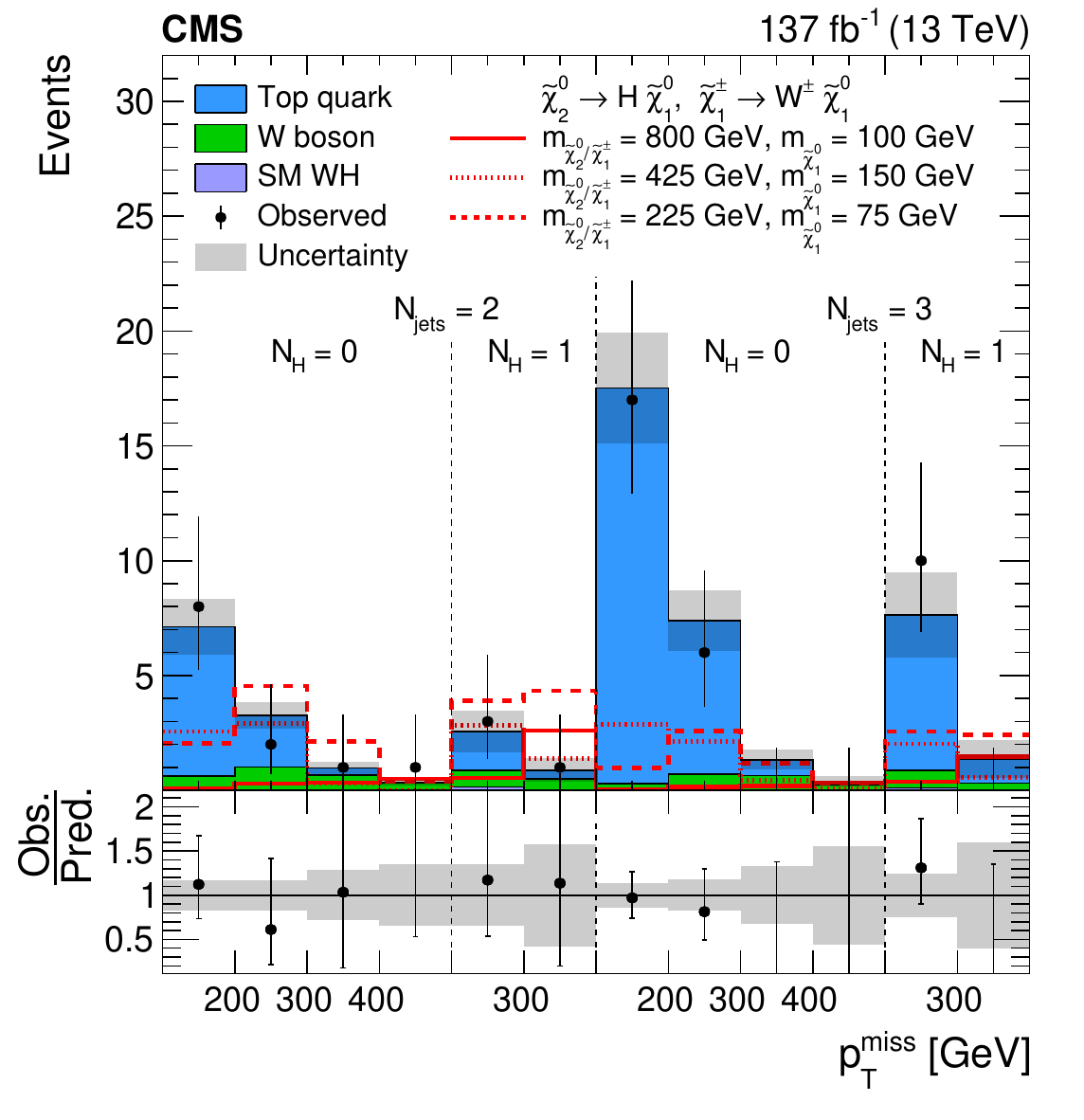}
\caption{
Predictions of the SM background after performing the signal extraction fit (filled histograms) and observed yields in the signal regions.
Three signal models with different values of $m_{\PSGczDt/\PSGcpmDo}$ and $m_{\PSGczDo}$ are shown as solid, short dashed, and long dashed lines.
The lower panel provides the ratio between the observation and the predicted SM backgrounds.
The shaded band shows the post-fit combination of the systematic and statistical uncertainties in the background prediction.
}
\label{fig:postFit}
\end{figure*}

The observed data yields and the expected yields from SM processes in the signal regions are summarized in Table~\ref{tab:summary_table}. No significant disagreement is observed.
A binned maximum likelihood fit for the SUSY signal strength, the yields of background events, and various nuisance parameters is performed.
The likelihood function is built using Poisson probability functions for all signal regions, and log-normal or gamma function PDFs for all nuisance parameters.
Figure~\ref{fig:postFit} shows the post-fit expectation of the SM background.
Combining all signal bins, $51 \pm 5$ background events are expected and 49 events are observed.

\begin{table}[bh]
  \centering

  \topcaption{
    Summary of the predicted SM background and the observed yield in the signal regions, together with the expected yields for three signal benchmark models. The total prediction, $N_{\mathrm{SR}}^{\mathrm{BG}}$, is the sum of the top quark and \PW boson predictions, $N_{\mathrm{SR}}^{\text{top}}$ and $N_{\mathrm{SR}}^{\mathrm{\PW}}$, as well as small contributions from standard model \PW{}\PH{} production. The values shown are taken before the signal extraction fit to the observed yields in the signal regions is performed. The uncertainties include the statistical and systematic components. For each benchmark model column, the ordered pairs indicate the masses (in GeV) of the \PSGczDt/\PSGcpmDo and \PSGczDo, respectively.}
  \label{tab:summary_table}
  \resizebox{1.0\textwidth}{!}{
  \begin{tabular}{cccccccccc}

\multirow{2}{*}{\njets} & \multirow{2}{*}{\nhiggs} & \multirow{2}{*}{\ptmiss [GeV]}& \multirow{2}{*}{$N_{\mathrm{SR}}^{\text{top}}$}& \multirow{2}{*}{$N_{\mathrm{SR}}^{\mathrm{\PW}}$} & \multirow{2}{*}{$N_{\mathrm{SR}}^{\mathrm{BG}}$} & \multirow{2}{*}{Observed} & \multicolumn{3}{c}{$\PSGczDt \to \PH \PSGczDo,~\PSGcpmDo \to \PW^{\pm} \PSGczDo$} \\
                   &                    &           &       & &           &      & 800, 100         & 425, 150       & 225, 75         \\ \hline
\multirow{6}{*}{2} & \multirow{4}{*}{0} &  125--200 & 6.3 &0.5&   $6.9 \pm 1.3$ &    8 &  $0.08 \pm 0.02$ &    $2.0 \pm 0.4$ &    $2.6 \pm 0.8$\\
                   &                    &  200--300 & 2.4 &1.0&   $3.4 \pm 0.6$ &    2 &  $0.3 \pm 0.1$ &    $4.5 \pm 0.7$ &    $2.9 \pm 0.6$\\
                   &                    &  300--400 & 0.3 &0.7&   $1.0 \pm 0.3$ &    1 &  $0.3 \pm 0.1$ &    $2.1 \pm 0.4$ &  $0.3 \pm 0.2$\\
                   &                    &    $>$400 & 0.02 &0.3&  $0.3 \pm 0.1$ &    1 &  $0.5 \pm 0.2$ &  $0.4 \pm 0.3$ &      $\leq 0.01$\\ [1\cmsTabSkip]
                   & \multirow{2}{*}{1} &  125--300 & 1.6 &0.7&   $2.5 \pm 0.9$ &    3 &  $0.5 \pm 0.1$ &    $3.9 \pm 0.7$ &    $2.8 \pm 1.0$\\
                   &                    &    $>$300 & 0.4 &0.5&   $0.9 \pm 0.5$ &    1 &   $2.6 \pm 0.4$ &    $4.3 \pm 0.8$ &    $1.4 \pm 0.4$\\ [2\cmsTabSkip]
\multirow{6}{*}{3} & \multirow{4}{*}{0} &  125--200 & 17.5 &0.2&  $17.8 \pm 3.0$ &   17 &  $0.05 \pm 0.02$ &  $1.0 \pm 0.2$ &    $2.9 \pm 0.6$\\
                   &                    &  200--300 & 7.1 &0.7&   $7.8 \pm 1.7$ &    6 &  $0.14 \pm 0.03$ &    $2.6 \pm 0.3$ &    $2.1 \pm 0.5$\\
                   &                    &  300--400 & 0.8 &0.7&   $1.5 \pm 0.5$ &    0 &  $0.18 \pm 0.04$ &    $1.2 \pm 0.4$ &  $0.4 \pm 0.4$\\
                   &                    &    $>$400 & 0.2 &0.3&   $0.5 \pm 0.3$ &    0 &  $0.3 \pm 0.1$ &  $0.3 \pm 0.2$ &  $0.06 \pm 0.06$\\ [1\cmsTabSkip]
                   & \multirow{2}{*}{1} &  125--300 & 5.0 &0.8&   $5.9 \pm 2.1$ &   10 &  $0.4 \pm 0.1$ &    $2.6 \pm 0.5$ &    $2.0 \pm 0.6$\\ 
                   &                    &    $>$300 & 1.7 &0.3&   $2.1 \pm 1.6$ &    0 &    $1.5 \pm 0.2$ &    $2.4 \pm 0.5$ &  $0.6 \pm 0.2$\\

\end{tabular}}
\end{table}

We next evaluate the experimental and theoretical uncertainties in the expected signal yield.
Varying the lepton, \PQb tagging, and \PH tagging efficiency scale factors by their respective uncertainties varies the signal yield by less than 1, 4, and 20\%.
For the \PH tagger, this scale factor is measured as a function of the \PH candidate \pt using a sample of jets in data and simulation that mimic the rare $\PH \to \PQb \PAQb$ case~\cite{Sirunyan:2020lcu}.

The efficiencies obtained using the fast or full detector simulation are found to be compatible, with no significant dependence on the mass splitting $\Delta m = m_{\PSGczDt/\PSGcpmDo}-m_{\PSGczDo}$.
The systematic uncertainty in the signal yields, due to the uncertainty in the trigger efficiency measurement, is generally less than 5\%.

The uncertainties in the simulated yields obtained by varying the jet energy scale and the jet energy resolution are each between 1 and 7\%. A 3\% difference in the \PQb jet energy scale between the fast and full detector simulations is observed,
resulting in a 1--10\% change in the expected signal yield.

The effect of missing higher-order corrections on the signal acceptance is estimated by varying $\mu_\mathrm{R}$ and $\mu_\mathrm{F}$~\cite{Catani:2003zt,Cacciari:2003fi,Kalogeropoulos:2018cke} up and down by a factor of 2, omitting the combination of variations in opposite directions.
The envelope of the variations reaches values up to 15\% and is assigned as a systematic uncertainty. The resulting variation of the expected signal yield is less than 1\%. 
To account for uncertainty in the modeling of the multiplicity of additional jets from initial state radiation, a 1\% uncertainty is applied to the $\njets=3$ signal regions.

The integrated luminosities of the 2016, 2017, and 2018 data-taking periods are individually known with uncertainties in the 2.3--2.5\% range~\cite{CMS-PAS-LUM-17-001,CMS-PAS-LUM-17-004,CMS-PAS-LUM-18-002}, while the total Run~2 (2016--2018) integrated luminosity has an uncertainty of 1.8\%, the improvement in precision reflecting the (uncorrelated) time evolution of some systematic effects.
The signal samples are reweighted according to the distribution of the true number of interactions per bunch crossing.
The uncertainty in the total inelastic \pp cross section leads to
changes in the expected signal yield of less than 2\%. A summary of the systematic uncertainties in the signal yields is given in Table~\ref{tab:sigSys}.

\begin{table}[]
    \centering
    \topcaption{Sources and ranges of systematic uncertainties on the expected signal yields. The ranges reported reflect the magnitudes of the median 68\% of all impacts, considering the distribution of variations in all 12 signal regions and the full range of signal mass hypotheses used. When the lower bound is very close to 0, an upper bound is shown instead.}
    \begin{tabular}{lc}
        Source                                        & Typical values \\ \hline 
        Simulation statistical uncertainty                 & 1--10\%     \\
        Lepton efficiency                        &  $<$1\%      \\ 
        \PQb tagging efficiency                    &  $<$4\%      \\ 
        \PH tagging efficiency                    &  7--20\%    \\ 
        Trigger efficiency                        &  $<$5\%      \\ 
        Jet energy scale                        &  1--7\%       \\ 
        Jet energy resolution                &  1--7\%       \\ 
        \PQb jet energy scale                     &  1--10\%       \\ 
        $\mu_\mathrm{R}$ and $\mu_\mathrm{F}$ &  $<$1\%        \\ 
        Initial-state radiation                                    &  1\%       \\ 
        Integrated luminosity                                 & 1.8\%            \\ 
        Pileup                                       &  $<$2\%        \\
    \end{tabular}
    \label{tab:sigSys}
\end{table}

The results are interpreted in the context of the simplified SUSY model shown in Fig.~\ref{fig:fey}.
The chargino and second-lightest neutralino are assumed to have the same mass, and the branching fractions for the decays shown are taken to be 100\%. Wino-like cross sections are assumed.
Cross section limits as a function of the masses of the produced particles are set using a modified frequentist approach at 95\% confidence level (\CL), with the \CLs criterion and an asymptotic formulation \cite{Junk_1999, Read_2002,Cowan:2010js}.
All signal regions are considered simultaneously and correlations among uncertainties are included.

\begin{figure}[hbt]
  \begin{center}
  \includegraphics[width=0.85\linewidth]{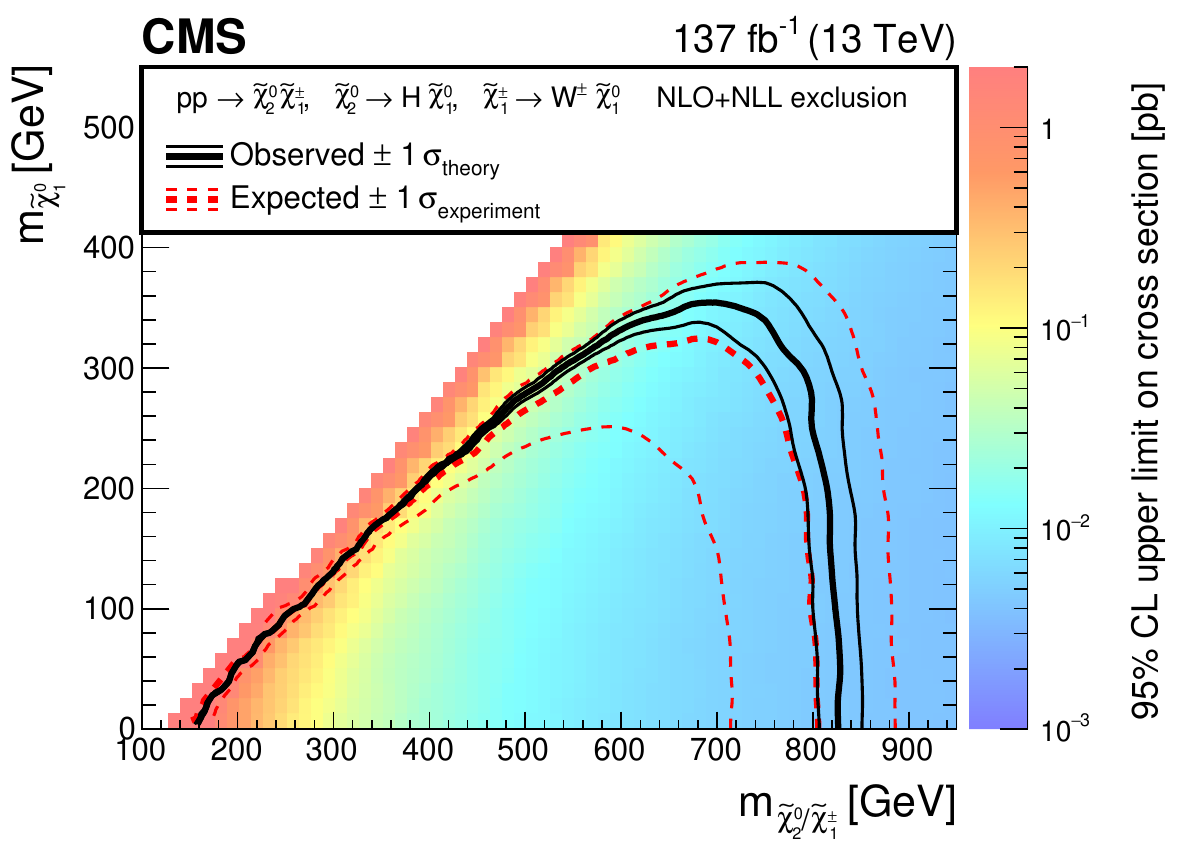}
  \caption{
     Cross section upper limits calculated with the background estimates and all of the background systematic uncertainties described in Sections \ref{ssec:topbg} and \ref{sec:wjetsbg}. 
     The color on the $z$ axis represents the 95\% CL upper limit on the cross section calculated at each point in the $m_{\PSGczDo}$-$m_{\PSGczDt}$ plane.
     The area below the thick black curve (dashed red line) represents the observed (expected) exclusion region at this \CL.
     The region containing 68\% of the distribution of limits expected under the background-only hypothesis is bounded by thin dashed red lines. The thin black lines show the effect of the theoretical uncertainties in the signal cross section.}
	\label{fig:limits}
      \end{center}
\end{figure}

Figure~\ref{fig:limits} shows the 95\% \CL upper limits on the cross section, together with the expected and observed exclusion limits in the $m_{\PSGczDo}$-$m_{\PSGczDt}$ plane for chargino-neutralino production. The effect of the uncertainty in the total production cross section due to the PDF model and the renormalization and refactorization scales is considered separately from the experimental uncertainties on the acceptance~\cite{PhysRevD.98.055014}, and is shown as the uncertainty band on the observed exclusion limits. 

This analysis excludes charginos with mass below 820\GeV for a low-mass LSP, and values of the LSP mass up to approximately 350\GeV for a chargino mass near 700\GeV. The excluded cross section for models with large mass splitting reaches approximately 5\unit{fb}.

\section{Summary}
\label{sec:summary}

This paper presents the results of a search for chargino-neutralino production in a final state containing a \PW boson decaying to leptons, a Higgs boson decaying to a bottom quark-antiquark pair, and missing transverse momentum. 
Expected yields from standard model processes are estimated by extrapolating the yields observed in control regions using transfer factors obtained from simulation.  
The observed yields agree with those expected from the standard model.
The results are interpreted as an exclusion of a simplified model of chargino-neutralino production. In the simplified model, the chargino decays to a \PW boson and a lightest supersymmetric particle (LSP), and the next-to-lightest neutralino decays to a Higgs boson and an LSP.
Charginos with mass below 820\GeV are excluded at 95\% confidence level for an LSP with mass below 200\GeV, and values of LSP mass up to approximately 350\GeV are excluded for a chargino mass near 700\GeV.

Relative to the previous result from the CMS Collaboration targeting this signature~\cite{Sirunyan:2017zss}, the sensitivity of the
search has been significantly extended.
The constraints on the masses of the chargino and LSP exceed those from the previous analysis by nearly 350 and 250\GeV, respectively. This represents a factor of 14 reduction in the excluded cross section for models with large mass splittings. Roughly half of this improvement is the result of the four-fold increase in integrated luminosity, with the remainder coming from analysis optimizations such as the inclusion of the \PH tagger and events with $\njets = 3$, as well as finer categorization of events based on \ptmiss made possible by the increased size of the data set.

\begin{acknowledgments}

  We congratulate our colleagues in the CERN accelerator departments for the excellent performance of the LHC and thank the technical and administrative staffs at CERN and at other CMS institutes for their contributions to the success of the CMS effort. In addition, we gratefully acknowledge the computing centers and personnel of the Worldwide LHC Computing Grid and other centers for delivering so effectively the computing infrastructure essential to our analyses. Finally, we acknowledge the enduring support for the construction and operation of the LHC, the CMS detector, and the supporting computing infrastructure provided by the following funding agencies: BMBWF and FWF (Austria); FNRS and FWO (Belgium); CNPq, CAPES, FAPERJ, FAPERGS, and FAPESP (Brazil); MES (Bulgaria); CERN; CAS, MoST, and NSFC (China); MINCIENCIAS (Colombia); MSES and CSF (Croatia); RIF (Cyprus); SENESCYT (Ecuador); MoER, ERC PUT and ERDF (Estonia); Academy of Finland, MEC, and HIP (Finland); CEA and CNRS/IN2P3 (France); BMBF, DFG, and HGF (Germany); GSRT (Greece); NKFIA (Hungary); DAE and DST (India); IPM (Iran); SFI (Ireland); INFN (Italy); MSIP and NRF (Republic of Korea); MES (Latvia); LAS (Lithuania); MOE and UM (Malaysia); BUAP, CINVESTAV, CONACYT, LNS, SEP, and UASLP-FAI (Mexico); MOS (Montenegro); MBIE (New Zealand); PAEC (Pakistan); MSHE and NSC (Poland); FCT (Portugal); JINR (Dubna); MON, RosAtom, RAS, RFBR, and NRC KI (Russia); MESTD (Serbia); SEIDI, CPAN, PCTI, and FEDER (Spain); MOSTR (Sri Lanka); Swiss Funding Agencies (Switzerland); MST (Taipei); ThEPCenter, IPST, STAR, and NSTDA (Thailand); TUBITAK and TAEK (Turkey); NASU (Ukraine); STFC (United Kingdom); DOE and NSF (USA).

  \hyphenation{Rachada-pisek} Individuals have received support from the Marie-Curie program and the European Research Council and Horizon 2020 Grant, contract Nos.\ 675440, 724704, 752730, 758316, 765710, 824093, and COST Action CA16108 (European Union); the Leventis Foundation; the Alfred P.\ Sloan Foundation; the Alexander von Humboldt Foundation; the Belgian Federal Science Policy Office; the Fonds pour la Formation \`a la Recherche dans l'Industrie et dans l'Agriculture (FRIA-Belgium); the Agentschap voor Innovatie door Wetenschap en Technologie (IWT-Belgium); the F.R.S.-FNRS and FWO (Belgium) under the ``Excellence of Science -- EOS" -- be.h project n.\ 30820817; the Beijing Municipal Science \& Technology Commission, No. Z191100007219010; the Ministry of Education, Youth and Sports (MEYS) of the Czech Republic; the Deutsche Forschungsgemeinschaft (DFG), under Germany's Excellence Strategy -- EXC 2121 ``Quantum Universe" -- 390833306, and under project number 400140256 - GRK2497; the Lend\"ulet (``Momentum") Program and the J\'anos Bolyai Research Scholarship of the Hungarian Academy of Sciences, the New National Excellence Program \'UNKP, the NKFIA research grants 123842, 123959, 124845, 124850, 125105, 128713, 128786, and 129058 (Hungary); the Council of Science and Industrial Research, India; the Latvian Council of Science; the Ministry of Science and Higher Education and the National Science Center, contracts Opus 2014/15/B/ST2/03998 and 2015/19/B/ST2/02861 (Poland); the National Priorities Research Program by Qatar National Research Fund; the Ministry of Science and Higher Education, project no. 0723-2020-0041 (Russia); the Programa Estatal de Fomento de la Investigaci{\'o}n Cient{\'i}fica y T{\'e}cnica de Excelencia Mar\'{\i}a de Maeztu, grant MDM-2015-0509 and the Programa Severo Ochoa del Principado de Asturias; the Stavros Niarchos Foundation (Greece); the Rachadapisek Sompot Fund for Postdoctoral Fellowship, Chulalongkorn University and the Chulalongkorn Academic into Its 2nd Century Project Advancement Project (Thailand); the Kavli Foundation; the Nvidia Corporation; the SuperMicro Corporation; the Welch Foundation, contract C-1845; and the Weston Havens Foundation (USA).
\end{acknowledgments}

\bibliography{auto_generated}  

\cleardoublepage \appendix\section{The CMS Collaboration \label{app:collab}}\begin{sloppypar}\hyphenpenalty=5000\widowpenalty=500\clubpenalty=5000\input{SUS-20-003-authorlist.tex}\end{sloppypar}
%%% END EDITABLE REGION %%%
% skeleton_end
\end{document}

%% file: SUS-20-003-authorlist.tex
\vskip\cmsinstskip
\textbf{Yerevan Physics Institute, Yerevan, Armenia}\\*[0pt]
A.~Tumasyan
\vskip\cmsinstskip
\textbf{Institut f\"{u}r Hochenergiephysik, Wien, Austria}\\*[0pt]
W.~Adam, J.W.~Andrejkovic, T.~Bergauer, S.~Chatterjee, M.~Dragicevic, A.~Escalante~Del~Valle, R.~Fr\"{u}hwirth\cmsAuthorMark{1}, M.~Jeitler\cmsAuthorMark{1}, N.~Krammer, L.~Lechner, D.~Liko, I.~Mikulec, P.~Paulitsch, F.M.~Pitters, J.~Schieck\cmsAuthorMark{1}, R.~Sch\"{o}fbeck, M.~Spanring, S.~Templ, W.~Waltenberger, C.-E.~Wulz\cmsAuthorMark{1}
% ...
\vskip\cmsinstskip
% ...
1:  Also at TU Wien, Wien, Austria\\
% ...


%% file: SUS-20-003_temp.bbl
\providecommand{\href}[2]{#2}\begingroup\raggedright\begin{thebibliography}{10}%
\makeatletter
\providecommand{\hrefCMSnoop }[0]{\@secondoftwo}%
\makeatother
\providecommand{\doi}{\texttt{doi:}\begingroup \urlstyle{tt}\Url}

\bibitem{Ramond:1971gb}
\hrefCMSnoop {}{P.~Ramond, ``{Dual theory for free fermions}'',} \textit{ Phys.
  Rev. D} \textbf{ 3} (1971) 2415,
  \href{http://dx.doi.org/10.1103/PhysRevD.3.2415}{\doi{10.1103/PhysRevD.3.2415}}.

\bibitem{Wess:1974tw}
\hrefCMSnoop {}{J.~Wess and B.~Zumino, ``{Supergauge transformations in four
  dimensions}'',} \textit{ Nucl. Phys. B} \textbf{ 70} (1974) 39,
  \href{http://dx.doi.org/10.1016/0550-3213(74)90355-1}{\doi{10.1016/0550-3213(74)90355-1}}.

\bibitem{Nilles:1983ge}
\hrefCMSnoop {}{H.~P. Nilles, ``{Supersymmetry, supergravity and particle
  physics}'',} \textit{ Phys. Rept.} \textbf{ 110} (1984) 1,
  \href{http://dx.doi.org/10.1016/0370-1573(84)90008-5}{\doi{10.1016/0370-1573(84)90008-5}}.

\bibitem{deCarlos:1993rbr}
\hrefCMSnoop {}{B.~de~Carlos and J.~A. Casas, ``{One-loop analysis of the
  electroweak breaking in supersymmetric models and the fine tuning
  problem}'',} \textit{ Phys. Lett. B} \textbf{ 309} (1993) 320,
  \href{http://dx.doi.org/10.1016/0370-2693(93)90940-J}{\doi{10.1016/0370-2693(93)90940-J}},
  \href{http://www.arXiv.org/abs/hep-ph/9303291}{\texttt{arXiv:hep-ph/9303291}}.

\bibitem{Dimopoulos:1995mi}
\hrefCMSnoop {}{S.~Dimopoulos and G.~F. Giudice, ``{Naturalness constraints in
  supersymmetric theories with nonuniversal soft terms}'',} \textit{ Phys.
  Lett. B} \textbf{ 357} (1995) 573,
  \href{http://dx.doi.org/10.1016/0370-2693(95)00961-J}{\doi{10.1016/0370-2693(95)00961-J}},
  \href{http://www.arXiv.org/abs/hep-ph/9507282}{\texttt{arXiv:hep-ph/9507282}}.

\bibitem{Farrar:1978xj}
\hrefCMSnoop {}{G.~R. Farrar and P.~Fayet, ``{Phenomenology of the production,
  decay, and detection of new hadronic states associated with
  supersymmetry}'',} \textit{ Phys. Lett. B} \textbf{ 76} (1978) 575,
  \href{http://dx.doi.org/10.1016/0370-2693(78)90858-4}{\doi{10.1016/0370-2693(78)90858-4}}.

\bibitem{Alwall:2008ag}
\hrefCMSnoop {}{J.~Alwall, P.~Schuster, and N.~Toro, ``Simplified models for a
  first characterization of new physics at the {LHC}'',} \textit{ Phys. Rev. D}
  \textbf{ 79} (2009) 075020,
  \href{http://dx.doi.org/10.1103/PhysRevD.79.075020}{\doi{10.1103/PhysRevD.79.075020}},
\href{http://www.arXiv.org/abs/0810.3921}{\texttt{arXiv:0810.3921}}.
%%CITATION = ARXIV:0810.3921;%%.

\bibitem{Alves:2011wf}
\hrefCMSnoop {}{{LHC New Physics Working Group}, ``Simplified models for {LHC}
  new physics searches'',} \textit{ J. Phys. G} \textbf{ 39} (2012) 105005,
  \href{http://dx.doi.org/10.1088/0954-3899/39/10/105005}{\doi{10.1088/0954-3899/39/10/105005}},
\href{http://www.arXiv.org/abs/1105.2838}{\texttt{arXiv:1105.2838}}.
%%CITATION = ARXIV:1105.2838;%%.

\bibitem{Chatrchyan:2013sza}
\hrefCMSnoop {}{{CMS Collaboration}, ``Interpretation of searches for
  supersymmetry with simplified models'',} \textit{ Phys. Rev. D} \textbf{ 88}
  (2013) 052017,
  \href{http://dx.doi.org/10.1103/PhysRevD.88.052017}{\doi{10.1103/PhysRevD.88.052017}},
\href{http://www.arXiv.org/abs/1301.2175}{\texttt{arXiv:1301.2175}}.
%%CITATION = ARXIV:1301.2175;%%.

\bibitem{ATLAS_WH_run2}
\hrefCMSnoop {}{{ATLAS Collaboration}, ``{Search for direct production of
  electroweakinos in final states with one lepton, missing transverse momentum
  and a Higgs boson decaying into two b-jets in pp collisions $\sqrt{s}=$ 13
  {TeV} with the ATLAS detector}'',} \textit{ Eur. Phys. J. C} \textbf{ 80}
  (2020) 691,
  \href{http://dx.doi.org/10.1140/epjc/s10052-020-8050-3}{\doi{10.1140/epjc/s10052-020-8050-3}},
  \href{http://www.arXiv.org/abs/1909.09226}{\texttt{arXiv:1909.09226}}.

\bibitem{Aad:2015eda}
\hrefCMSnoop {}{{ATLAS Collaboration}, ``{Search for the electroweak production
  of supersymmetric particles in $\sqrt{s}$ = 8 TeV $pp$ collisions with the
  ATLAS detector}'',} \textit{ Phys. Rev. D} \textbf{ 93} (2016) 052002,
  \href{http://dx.doi.org/10.1103/PhysRevD.93.052002}{\doi{10.1103/PhysRevD.93.052002}},
  \href{http://www.arXiv.org/abs/1509.07152}{\texttt{arXiv:1509.07152}}.

\bibitem{Sirunyan:2017zss}
\hrefCMSnoop {}{{CMS Collaboration}, ``{Search for electroweak production of
  charginos and neutralinos in WH events in proton-proton collisions at $
  \sqrt{s}=13 $ TeV}'',} \textit{ JHEP} \textbf{ 11} (2017) 029,
  \href{http://dx.doi.org/10.1007/JHEP11(2017)029}{\doi{10.1007/JHEP11(2017)029}},
  \href{http://www.arXiv.org/abs/1706.09933}{\texttt{arXiv:1706.09933}}.

\bibitem{Khachatryan:2014mma}
\hrefCMSnoop {}{{CMS Collaboration}, ``{Searches for electroweak neutralino and
  chargino production in channels with Higgs, Z, and W bosons in pp collisions
  at 8 TeV}'',} \textit{ Phys. Rev. D} \textbf{ 90} (2014) 092007,
  \href{http://dx.doi.org/10.1103/PhysRevD.90.092007}{\doi{10.1103/PhysRevD.90.092007}},
  \href{http://www.arXiv.org/abs/1409.3168}{\texttt{arXiv:1409.3168}}.

\bibitem{Khachatryan:2014qwa}
\hrefCMSnoop {}{{CMS Collaboration}, ``{Searches for electroweak production of
  charginos, neutralinos, and sleptons decaying to leptons and W, Z, and Higgs
  bosons in pp collisions at 8 TeV}'',} \textit{ Eur. Phys. J. C} \textbf{ 74}
  (2014) 3036,
  \href{http://dx.doi.org/10.1140/epjc/s10052-014-3036-7}{\doi{10.1140/epjc/s10052-014-3036-7}},
  \href{http://www.arXiv.org/abs/1405.7570}{\texttt{arXiv:1405.7570}}.

\bibitem{Chatrchyan:2008zzk}
\hrefCMSnoop {}{{CMS Collaboration}, ``The {CMS} experiment at the {CERN}
  {LHC}'',} \textit{ JINST} \textbf{ 3} (2008) S08004,
  \href{http://dx.doi.org/10.1088/1748-0221/3/08/S08004}{\doi{10.1088/1748-0221/3/08/S08004}}.

\bibitem{Sirunyan:2020zal}
\hrefCMSnoop {}{{CMS Collaboration}, ``{Performance of the CMS Level-1 trigger
  in proton-proton collisions at $\sqrt{s} = 13$\,TeV}'',} \textit{ JINST}
  \textbf{ 15} (2020) P10017,
  \href{http://dx.doi.org/10.1088/1748-0221/15/10/P10017}{\doi{10.1088/1748-0221/15/10/P10017}},
  \href{http://www.arXiv.org/abs/2006.10165}{\texttt{arXiv:2006.10165}}.

\bibitem{Khachatryan:2016bia}
\hrefCMSnoop {}{{CMS Collaboration}, ``{The CMS trigger system}'',} \textit{
  JINST} \textbf{ 12} (2017) P01020,
  \href{http://dx.doi.org/10.1088/1748-0221/12/01/P01020}{\doi{10.1088/1748-0221/12/01/P01020}},
\href{http://www.arXiv.org/abs/1609.02366}{\texttt{arXiv:1609.02366}}.
%%CITATION = ARXIV:1609.02366;%%.

\bibitem{Alwall:2014hca}
J.~Alwall\hrefCMSnoop {}{ {et~al.}, ``The automated computation of tree-level
  and next-to-leading order differential cross sections, and their matching to
  parton shower simulations'',} \textit{ JHEP} \textbf{ 07} (2014) 079,
  \href{http://dx.doi.org/10.1007/JHEP07(2014)079}{\doi{10.1007/JHEP07(2014)079}},
\href{http://www.arXiv.org/abs/1405.0301}{\texttt{arXiv:1405.0301}}.
%%CITATION = ARXIV:1405.0301;%%.

\bibitem{Nason:2004rx}
\hrefCMSnoop {}{P.~Nason, ``{A new method for combining NLO QCD with shower
  Monte Carlo algorithms}'',} \textit{ JHEP} \textbf{ 11} (2004) 040,
  \href{http://dx.doi.org/10.1088/1126-6708/2004/11/040}{\doi{10.1088/1126-6708/2004/11/040}},
\href{http://www.arXiv.org/abs/hep-ph/0409146}{\texttt{arXiv:hep-ph/0409146}}.
%%CITATION = HEP-PH/0409146;%%.

\bibitem{Frixione:2007vw}
\hrefCMSnoop {}{S.~Frixione, P.~Nason, and C.~Oleari, ``{Matching NLO QCD
  computations with parton shower simulations: the \POWHEG method}'',} \textit{
  JHEP} \textbf{ 11} (2007) 070,
  \href{http://dx.doi.org/10.1088/1126-6708/2007/11/070}{\doi{10.1088/1126-6708/2007/11/070}},
\href{http://www.arXiv.org/abs/0709.2092}{\texttt{arXiv:0709.2092}}.
%%CITATION = ARXIV:0709.2092;%%.

\bibitem{Alioli:2010xd}
\hrefCMSnoop {}{S.~Alioli, P.~Nason, C.~Oleari, and E.~Re, ``{A general
  framework for implementing NLO calculations in shower Monte Carlo programs:
  the \POWHEG BOX}'',} \textit{ JHEP} \textbf{ 06} (2010) 043,
  \href{http://dx.doi.org/10.1007/JHEP06(2010)043}{\doi{10.1007/JHEP06(2010)043}},
\href{http://www.arXiv.org/abs/1002.2581}{\texttt{arXiv:1002.2581}}.
%%CITATION = ARXIV:1002.2581;%%.

\bibitem{Re:2010bp}
\hrefCMSnoop {}{E.~Re, ``{Single-top $\PW\PQt$-channel production matched with
  parton showers using the POWHEG method}'',} \textit{ Eur. Phys. J. C}
  \textbf{ 71} (2011) 1547,
  \href{http://dx.doi.org/10.1140/epjc/s10052-011-1547-z}{\doi{10.1140/epjc/s10052-011-1547-z}},
\href{http://www.arXiv.org/abs/1009.2450}{\texttt{arXiv:1009.2450}}.
%%CITATION = ARXIV:1009.2450;%%.

\bibitem{Ball:2011uy}
\hrefCMSnoop {}{{NNPDF} Collaboration, ``{Unbiased global determination of
  parton distributions and their uncertainties at NNLO and at LO}'',} \textit{
  Nucl. Phys. B} \textbf{ 855} (2012) 153,
  \href{http://dx.doi.org/10.1016/j.nuclphysb.2011.09.024}{\doi{10.1016/j.nuclphysb.2011.09.024}},
\href{http://www.arXiv.org/abs/1107.2652}{\texttt{arXiv:1107.2652}}.
%%CITATION = ARXIV:1107.2652;%%.

\bibitem{Ball:2014uwa}
\hrefCMSnoop {}{{NNPDF} Collaboration, ``{Parton distributions for the LHC Run
  II}'',} \textit{ JHEP} \textbf{ 04} (2015) 040,
  \href{http://dx.doi.org/10.1007/JHEP04(2015)040}{\doi{10.1007/JHEP04(2015)040}},
\href{http://www.arXiv.org/abs/1410.8849}{\texttt{arXiv:1410.8849}}.
%%CITATION = ARXIV:1410.8849;%%.

\bibitem{Ball:2017nwa}
\hrefCMSnoop {}{{NNPDF} Collaboration, ``{Parton distributions from
  high-precision collider data}'',} \textit{ Eur. Phys. J. C} \textbf{ 77}
  (2017) 663,
  \href{http://dx.doi.org/10.1140/epjc/s10052-017-5199-5}{\doi{10.1140/epjc/s10052-017-5199-5}},
\href{http://www.arXiv.org/abs/1706.00428}{\texttt{arXiv:1706.00428}}.
%%CITATION = ARXIV:1706.00428;%%.

\bibitem{Sjostrand:2014zea}
T.~Sj{\"o}strand\hrefCMSnoop {}{ {et~al.}, ``An introduction to {PYTHIA}
  8.2'',} \textit{ Comput. Phys. Commun.} \textbf{ 191} (2015) 159,
  \href{http://dx.doi.org/10.1016/j.cpc.2015.01.024}{\doi{10.1016/j.cpc.2015.01.024}},
\href{http://www.arXiv.org/abs/1410.3012}{\texttt{arXiv:1410.3012}}.
%%CITATION = ARXIV:1410.3012;%%.

\bibitem{Alwall:2007fs}
J.~Alwall\hrefCMSnoop {}{ {et~al.}, ``Comparative study of various algorithms
  for the merging of parton showers and matrix elements in hadronic
  collisions'',} \textit{ Eur. Phys. J. C} \textbf{ 53} (2008) 473,
  \href{http://dx.doi.org/10.1140/epjc/s10052-007-0490-5}{\doi{10.1140/epjc/s10052-007-0490-5}},
\href{http://www.arXiv.org/abs/0706.2569}{\texttt{arXiv:0706.2569}}.
%%CITATION = ARXIV:0706.2569;%%.

\bibitem{Frederix:2012ps}
\hrefCMSnoop {}{R.~Frederix and S.~Frixione, ``{Merging meets matching in
  MC@NLO}'',} \textit{ JHEP} \textbf{ 12} (2012) 061,
  \href{http://dx.doi.org/10.1007/JHEP12(2012)061}{\doi{10.1007/JHEP12(2012)061}},
\href{http://www.arXiv.org/abs/1209.6215}{\texttt{arXiv:1209.6215}}.
%%CITATION = ARXIV:1209.6215;%%.

\bibitem{Khachatryan:2015pea}
\hrefCMSnoop {}{{CMS Collaboration}, ``{Event generator tunes obtained from
  underlying event and multiparton scattering measurements}'',} \textit{ Eur.
  Phys. J. C} \textbf{ 76} (2016) 155,
  \href{http://dx.doi.org/10.1140/epjc/s10052-016-3988-x}{\doi{10.1140/epjc/s10052-016-3988-x}},
\href{http://www.arXiv.org/abs/1512.00815}{\texttt{arXiv:1512.00815}}.
%%CITATION = ARXIV:1512.00815;%%.

\bibitem{Sirunyan:2019dfx}
\hrefCMSnoop {}{{CMS Collaboration}, ``{Extraction and validation of a new set
  of CMS PYTHIA8 tunes from underlying-event measurements}'',} \textit{ Eur.
  Phys. J. C} \textbf{ 80} (2020) 4,
  \href{http://dx.doi.org/10.1140/epjc/s10052-019-7499-4}{\doi{10.1140/epjc/s10052-019-7499-4}},
  \href{http://www.arXiv.org/abs/1903.12179}{\texttt{arXiv:1903.12179}}.

\bibitem{Agostinelli2003250}
\hrefCMSnoop {}{{\GEANTfour} Collaboration, ``{\GEANTfour}~---~a simulation
  toolkit'',} \textit{ Nucl. Instrum. Meth. A} \textbf{ 506} (2003) 250,
\href{http://dx.doi.org/10.1016/S0168-9002(03)01368-8}{\doi{10.1016/S0168-9002(03)01368-8}}.
%%CITATION = NUIMA,A506,250;%%.

\bibitem{Abdullin:2011zz}
S.~Abdullin\hrefCMSnoop {}{ {et~al.}, ``{The fast simulation of the CMS
  detector at LHC}'',} \textit{ J. Phys. Conf. Ser.} \textbf{ 331} (2011)
  032049,
\href{http://dx.doi.org/10.1088/1742-6596/331/3/032049}{\doi{10.1088/1742-6596/331/3/032049}}.
%%CITATION = 00462,331,032049;%%.

\bibitem{Giammanco:2014bza}
\hrefCMSnoop {}{A.~Giammanco, ``{The fast simulation of the CMS experiment}'',}
  \textit{ J. Phys. Conf. Ser.} \textbf{ 513} (2014) 022012,
\href{http://dx.doi.org/10.1088/1742-6596/513/2/022012}{\doi{10.1088/1742-6596/513/2/022012}}.
%%CITATION = 00462,513,022012;%%.

\bibitem{Li:2012wna}
\hrefCMSnoop {}{Y.~Li and F.~Petriello, ``{Combining QCD and electroweak
  corrections to dilepton production in FEWZ}'',} \textit{ Phys. Rev. D}
  \textbf{ 86} (2012) 094034,
  \href{http://dx.doi.org/10.1103/PhysRevD.86.094034}{\doi{10.1103/PhysRevD.86.094034}},
\href{http://www.arXiv.org/abs/1208.5967}{\texttt{arXiv:1208.5967}}.
%%CITATION = ARXIV:1208.5967;%%.

\bibitem{Aliev:2010zk}
M.~Aliev\hrefCMSnoop {}{ {et~al.}, ``{HATHOR: HAdronic Top and Heavy quarks
  crOss section calculatoR}'',} \textit{ Comput. Phys. Commun.} \textbf{ 182}
  (2011) 1034,
  \href{http://dx.doi.org/10.1016/j.cpc.2010.12.040}{\doi{10.1016/j.cpc.2010.12.040}},
\href{http://www.arXiv.org/abs/1007.1327}{\texttt{arXiv:1007.1327}}.
%%CITATION = ARXIV:1007.1327;%%.

\bibitem{Kant:2014oha}
P.~Kant\hrefCMSnoop {}{ {et~al.}, ``{HatHor for single top-quark production:
  Updated predictions and uncertainty estimates for single top-quark production
  in hadronic collisions}'',} \textit{ Comput. Phys. Commun.} \textbf{ 191}
  (2015) 74,
  \href{http://dx.doi.org/10.1016/j.cpc.2015.02.001}{\doi{10.1016/j.cpc.2015.02.001}},
\href{http://www.arXiv.org/abs/1406.4403}{\texttt{arXiv:1406.4403}}.
%%CITATION = ARXIV:1406.4403;%%.

\bibitem{Beneke:2011mq}
\hrefCMSnoop {}{M.~Beneke, P.~Falgari, S.~Klein, and C.~Schwinn, ``{Hadronic
  top-quark pair production with NNLL threshold resummation}'',} \textit{ Nucl.
  Phys. B} \textbf{ 855} (2012) 695,
  \href{http://dx.doi.org/10.1016/j.nuclphysb.2011.10.021}{\doi{10.1016/j.nuclphysb.2011.10.021}},
\href{http://www.arXiv.org/abs/1109.1536}{\texttt{arXiv:1109.1536}}.
%%CITATION = ARXIV:1109.1536;%%.

\bibitem{Cacciari:2011hy}
M.~Cacciari\hrefCMSnoop {}{ {et~al.}, ``{Top-pair production at hadron
  colliders with next-to-next-to-leading logarithmic soft-gluon
  resummation}'',} \textit{ Phys. Lett. B} \textbf{ 710} (2012) 612,
  \href{http://dx.doi.org/10.1016/j.physletb.2012.03.013}{\doi{10.1016/j.physletb.2012.03.013}},
\href{http://www.arXiv.org/abs/1111.5869}{\texttt{arXiv:1111.5869}}.
%%CITATION = ARXIV:1111.5869;%%.

\bibitem{Czakon:2011xx}
\hrefCMSnoop {}{M.~Czakon and A.~Mitov, ``{Top++: A program for the calculation
  of the top-pair cross-section at hadron colliders}'',} \textit{ Comput. Phys.
  Commun.} \textbf{ 185} (2014) 2930,
  \href{http://dx.doi.org/10.1016/j.cpc.2014.06.021}{\doi{10.1016/j.cpc.2014.06.021}},
\href{http://www.arXiv.org/abs/1112.5675}{\texttt{arXiv:1112.5675}}.
%%CITATION = ARXIV:1112.5675;%%.

\bibitem{Baernreuther:2012ws}
\hrefCMSnoop {}{P.~B{\"{a}}rnreuther, M.~Czakon, and A.~Mitov, ``{Percent level
  precision physics at the Tevatron: First genuine NNLO QCD corrections to $q
  \bar{q} \to t \bar{t} + X$}'',} \textit{ Phys. Rev. Lett.} \textbf{ 109}
  (2012) 132001,
  \href{http://dx.doi.org/10.1103/PhysRevLett.109.132001}{\doi{10.1103/PhysRevLett.109.132001}},
\href{http://www.arXiv.org/abs/1204.5201}{\texttt{arXiv:1204.5201}}.
%%CITATION = ARXIV:1204.5201;%%.

\bibitem{Czakon:2012zr}
\hrefCMSnoop {}{M.~Czakon and A.~Mitov, ``{NNLO corrections to top-pair
  production at hadron colliders: the all-fermionic scattering channels}'',}
  \textit{ JHEP} \textbf{ 12} (2012) 054,
  \href{http://dx.doi.org/10.1007/JHEP12(2012)054}{\doi{10.1007/JHEP12(2012)054}},
\href{http://www.arXiv.org/abs/1207.0236}{\texttt{arXiv:1207.0236}}.
%%CITATION = ARXIV:1207.0236;%%.

\bibitem{Czakon:2012pz}
\hrefCMSnoop {}{M.~Czakon and A.~Mitov, ``{NNLO corrections to top pair
  production at hadron colliders: the quark-gluon reaction}'',} \textit{ JHEP}
  \textbf{ 01} (2013) 080,
  \href{http://dx.doi.org/10.1007/JHEP01(2013)080}{\doi{10.1007/JHEP01(2013)080}},
\href{http://www.arXiv.org/abs/1210.6832}{\texttt{arXiv:1210.6832}}.
%%CITATION = ARXIV:1210.6832;%%.

\bibitem{Czakon:2013goa}
\hrefCMSnoop {}{M.~Czakon, P.~Fiedler, and A.~Mitov, ``{Total top-quark
  pair-production cross section at hadron colliders through $O(\alpS^4)$}'',}
  \textit{ Phys. Rev. Lett.} \textbf{ 110} (2013) 252004,
  \href{http://dx.doi.org/10.1103/PhysRevLett.110.252004}{\doi{10.1103/PhysRevLett.110.252004}},
\href{http://www.arXiv.org/abs/1303.6254}{\texttt{arXiv:1303.6254}}.
%%CITATION = ARXIV:1303.6254;%%.

\bibitem{PhysRevLett.83.3780}
W.~Beenakker\hrefCMSnoop {}{ {et~al.}, ``Production of charginos, neutralinos,
  and sleptons at hadron colliders'',} \textit{ Phys. Rev. Lett.} \textbf{ 83}
  (1999) 3780,
  \href{http://dx.doi.org/10.1103/PhysRevLett.83.3780}{\doi{10.1103/PhysRevLett.83.3780}},
  \href{http://www.arXiv.org/abs/hep-ph/9906298}{\texttt{arXiv:hep-ph/9906298}}.

\bibitem{DEBOVE201151}
\hrefCMSnoop {}{J.~Debove, B.~Fuks, and M.~Klasen, ``Threshold resummation for
  gaugino pair production at hadron colliders'',} \textit{ Nuc. Phys. B}
  \textbf{ 842} (2011)
  \href{http://dx.doi.org/10.1016/j.nuclphysb.2010.08.016}{\doi{10.1016/j.nuclphysb.2010.08.016}},
  \href{http://www.arXiv.org/abs/1005.2909}{\texttt{arXiv:1005.2909}}.

\bibitem{resummino}
\hrefCMSnoop {}{B.~Fuks, M.~Klasen, D.~R. Lamprea, and M.~Rothering,
  ``Precision predictions for electroweak superpartner production at hadron
  colliders with resummino'',} \textit{ Eur. Phys. J. C} \textbf{ 73} (2013)
  2480,
  \href{http://dx.doi.org/10.1140/epjc/s10052-013-2480-0}{\doi{10.1140/epjc/s10052-013-2480-0}},
  \href{http://www.arXiv.org/abs/1304.0790}{\texttt{arXiv:1304.0790}}.

\bibitem{PhysRevD.98.055014}
\hrefCMSnoop {}{J.~Fiaschi and M.~Klasen, ``{Neutralino-chargino pair
  production at $\mathrm{NLO}\text{+NLL}$ with resummation-improved parton
  density functions for LHC Run II}'',} \textit{ Phys. Rev. D} \textbf{ 98}
  (2018) 055014,
  \href{http://dx.doi.org/10.1103/PhysRevD.98.055014}{\doi{10.1103/PhysRevD.98.055014}},
  \href{http://www.arXiv.org/abs/1805.11322}{\texttt{arXiv:1805.11322}}.

\bibitem{deFlorian:2016spz}
\hrefCMSnoop {}{{LHC Higgs Cross Section Working Group}, ``Handbook of {LHC}
  {H}iggs cross sections: 4. {D}eciphering the nature of the {H}iggs sector'',}
  CERN Report CERN-2017-002-M, 2016.
\newblock
  \href{http://dx.doi.org/10.23731/CYRM-2017-002}{\doi{10.23731/CYRM-2017-002}},
  \href{http://www.arXiv.org/abs/1610.07922}{\texttt{arXiv:1610.07922}}.

\bibitem{Sirunyan_2020_032}
\hrefCMSnoop {}{{CMS Collaboration}, ``Search for direct top squark pair
  production in events with one lepton, jets, and missing transverse momentum
  at 13 {TeV} with the {CMS} experiment'',} \textit{ JHEP} \textbf{ 05} (2020)
  032,
  \href{http://dx.doi.org/10.1007/jhep05(2020)032}{\doi{10.1007/jhep05(2020)032}},
  \href{http://www.arXiv.org/abs/1912.08887}{\texttt{arXiv:1912.08887}}.

\bibitem{Sirunyan:2017ulk}
\hrefCMSnoop {}{{CMS Collaboration}, ``Particle-flow reconstruction and global
  event description with the {CMS} detector'',} \textit{ JINST} \textbf{ 12}
  (2017) P10003,
  \href{http://dx.doi.org/10.1088/1748-0221/12/10/P10003}{\doi{10.1088/1748-0221/12/10/P10003}},
\href{http://www.arXiv.org/abs/1706.04965}{\texttt{arXiv:1706.04965}}.
%%CITATION = ARXIV:1706.04965;%%.

\bibitem{Sirunyan:2019kia}
\hrefCMSnoop {}{{CMS Collaboration}, ``{Performance of missing transverse
  momentum reconstruction in proton-proton collisions at $\sqrt{s} =$ 13 TeV
  using the CMS detector}'',} \textit{ JINST} \textbf{ 14} (2019) P07004,
  \href{http://dx.doi.org/10.1088/1748-0221/14/07/P07004}{\doi{10.1088/1748-0221/14/07/P07004}},
\href{http://www.arXiv.org/abs/1903.06078}{\texttt{arXiv:1903.06078}}.
%%CITATION = ARXIV:1903.06078;%%.

\bibitem{Sirunyan:2020ycc}
\hrefCMSnoop {}{{CMS Collaboration}, ``Electron and photon reconstruction and
  identification with the {CMS} experiment at the {CERN} {LHC}'',} \textit{
  JINST} \textbf{ 16} (2021) P05014,
  \href{http://dx.doi.org/10.1088/1748-0221/16/05/p05014}{\doi{10.1088/1748-0221/16/05/p05014}},
  \href{http://www.arXiv.org/abs/2012.06888}{\texttt{arXiv:2012.06888}}.

\bibitem{Sirunyan:2018fpa}
\hrefCMSnoop {}{{CMS Collaboration}, ``Performance of the {CMS} muon detector
  and muon reconstruction with proton-proton collisions at
  {$\sqrt{s}=13\TeV$}'',} \textit{ JINST} \textbf{ 13} (2018) P06015,
  \href{http://dx.doi.org/10.1088/1748-0221/13/06/P06015}{\doi{10.1088/1748-0221/13/06/P06015}},
\href{http://www.arXiv.org/abs/1804.04528}{\texttt{arXiv:1804.04528}}.
%%CITATION = ARXIV:1804.04528;%%.

\bibitem{Khachatryan:2015dfa}
\hrefCMSnoop {}{{CMS Collaboration}, ``{Reconstruction and identification of
  \ensuremath{\tau} lepton decays to hadrons and \ensuremath{\nu}$_{\tau}$ at
  CMS}'',} \textit{ JINST} \textbf{ 11} (2016) P01019,
  \href{http://dx.doi.org/10.1088/1748-0221/11/01/P01019}{\doi{10.1088/1748-0221/11/01/P01019}},
  \href{http://www.arXiv.org/abs/1510.07488}{\texttt{arXiv:1510.07488}}.

\bibitem{CMS:2016gvn}
\hrefCMSnoop {}{{CMS Collaboration}, ``Performance of reconstruction and
  identification of $\uptau$ leptons decaying to hadrons and
  \ensuremath{\nu}$_{\tau}$ in pp collisions at $\sqrt{s} = 13$ {TeV}'',}
  \textit{ JINST} \textbf{ 13} (2018) P10005,
  \href{http://dx.doi.org/10.1088/1748-0221/13/10/p10005}{\doi{10.1088/1748-0221/13/10/p10005}},
  \href{http://www.arXiv.org/abs/1809.02816}{\texttt{arXiv:1809.02816}}.

\bibitem{Cacciari:2008gp}
\hrefCMSnoop {}{M.~Cacciari, G.~P. Salam, and G.~Soyez, ``The anti-{$\kt$} jet
  clustering algorithm'',} \textit{ JHEP} \textbf{ 04} (2008) 063,
  \href{http://dx.doi.org/10.1088/1126-6708/2008/04/063}{\doi{10.1088/1126-6708/2008/04/063}},
  \href{http://www.arXiv.org/abs/0802.1189}{\texttt{arXiv:0802.1189}}.

\bibitem{Cacciari:2011ma}
\hrefCMSnoop {}{M.~Cacciari, G.~P. Salam, and G.~Soyez, ``{FastJet} user
  manual'',} \textit{ Eur. Phys. J. C} \textbf{ 72} (2012) 1896,
  \href{http://dx.doi.org/10.1140/epjc/s10052-012-1896-2}{\doi{10.1140/epjc/s10052-012-1896-2}},
\href{http://www.arXiv.org/abs/1111.6097}{\texttt{arXiv:1111.6097}}.
%%CITATION = ARXIV:1111.6097;%%.

\bibitem{Chatrchyan:2011ds}
\hrefCMSnoop {}{{CMS Collaboration}, ``{Determination of jet energy calibration
  and transverse momentum resolution in CMS}'',} \textit{ JINST} \textbf{ 6}
  (2011) P11002,
  \href{http://dx.doi.org/10.1088/1748-0221/6/11/P11002}{\doi{10.1088/1748-0221/6/11/P11002}},
  \href{http://www.arXiv.org/abs/1107.4277}{\texttt{arXiv:1107.4277}}.

\bibitem{Sirunyan:2017ezt}
\hrefCMSnoop {}{{CMS Collaboration}, ``{Identification of heavy-flavour jets
  with the CMS detector in $\Pp\Pp$ collisions at 13 TeV}'',} \textit{ JINST}
  \textbf{ 13} (2018) P05011,
  \href{http://dx.doi.org/10.1088/1748-0221/13/05/P05011}{\doi{10.1088/1748-0221/13/05/P05011}},
\href{http://www.arXiv.org/abs/1712.07158}{\texttt{arXiv:1712.07158}}.
%%CITATION = ARXIV:1712.07158;%%.

\bibitem{Sirunyan:2020lcu}
\hrefCMSnoop {}{{CMS Collaboration}, ``{Identification of heavy, energetic,
  hadronically decaying particles using machine-learning techniques}'',}
  \textit{ JINST} \textbf{ 15} (2020) P06005,
  \href{http://dx.doi.org/10.1088/1748-0221/15/06/P06005}{\doi{10.1088/1748-0221/15/06/P06005}},
  \href{http://www.arXiv.org/abs/2004.08262}{\texttt{arXiv:2004.08262}}.

\bibitem{Chatrchyan:2011tn}
\hrefCMSnoop {}{{CMS Collaboration}, ``{Missing transverse energy performance
  of the CMS detector}'',} \textit{ JINST} \textbf{ 6} (2011) P09001,
  \href{http://dx.doi.org/10.1088/1748-0221/6/09/P09001}{\doi{10.1088/1748-0221/6/09/P09001}},
\href{http://www.arXiv.org/abs/1106.5048}{\texttt{arXiv:1106.5048}}.
%%CITATION = ARXIV:1106.5048;%%.

\bibitem{Tovey:2008ui}
\hrefCMSnoop {}{D.~R. Tovey, ``{On measuring the masses of pair-produced
  semi-invisibly decaying particles at hadron colliders}'',} \textit{ JHEP}
  \textbf{ 04} (2008) 034,
  \href{http://dx.doi.org/10.1088/1126-6708/2008/04/034}{\doi{10.1088/1126-6708/2008/04/034}},
\href{http://www.arXiv.org/abs/0802.2879}{\texttt{arXiv:0802.2879}}.
%%CITATION = ARXIV:0802.2879;%%.

\bibitem{Sirunyan:2018kst}
\hrefCMSnoop {}{{CMS Collaboration}, ``{Observation of Higgs boson decay to
  bottom quarks}'',} \textit{ Phys. Rev. Lett.} \textbf{ 121} (2018) 121801,
  \href{http://dx.doi.org/10.1103/PhysRevLett.121.121801}{\doi{10.1103/PhysRevLett.121.121801}},
  \href{http://www.arXiv.org/abs/1808.08242}{\texttt{arXiv:1808.08242}}.

\bibitem{Sirunyan:2019bez}
\hrefCMSnoop {}{{CMS Collaboration}, ``{Measurements of the pp $\to$ WZ
  inclusive and differential production cross section and constraints on
  charged anomalous triple gauge couplings at $\sqrt{s} =$ 13 TeV}'',} \textit{
  JHEP} \textbf{ 04} (2019) 122,
  \href{http://dx.doi.org/10.1007/JHEP04(2019)122}{\doi{10.1007/JHEP04(2019)122}},
  \href{http://www.arXiv.org/abs/1901.03428}{\texttt{arXiv:1901.03428}}.

\bibitem{Sirunyan:2020jtq}
\hrefCMSnoop {}{{CMS Collaboration}, ``{W$^+$W$^-$ boson pair production in
  proton-proton collisions at $\sqrt{s} =$ 13 TeV}'',} \textit{ Phys. Rev. D}
  \textbf{ 102} (2020) 092001,
  \href{http://dx.doi.org/10.1103/PhysRevD.102.092001}{\doi{10.1103/PhysRevD.102.092001}},
  \href{http://www.arXiv.org/abs/2009.00119}{\texttt{arXiv:2009.00119}}.

\bibitem{Sirunyan:2020pub}
\hrefCMSnoop {}{{CMS Collaboration}, ``{Measurements of ${\mathrm{p}}
  {\mathrm{p}} \rightarrow {\mathrm{Z}} {\mathrm{Z}} $ production cross
  sections and constraints on anomalous triple gauge couplings at $\sqrt{s} =
  13\,\text {TeV} $}'',} \textit{ Eur. Phys. J. C} \textbf{ 81} (2021) 200,
  \href{http://dx.doi.org/10.1140/epjc/s10052-020-08817-8}{\doi{10.1140/epjc/s10052-020-08817-8}},
  \href{http://www.arXiv.org/abs/2009.01186}{\texttt{arXiv:2009.01186}}.

\bibitem{CMS-DP-2020-019}
\href {https://cds.cern.ch/record/2715872}{{CMS Collaboration}, ``{Jet energy
  scale and resolution performance with 13 TeV data collected by CMS in
  2016-2018}'',} {CMS Detector Performance Summary,} CMS-DP-2020-019, 2020.

\bibitem{Catani:2003zt}
\hrefCMSnoop {}{S.~Catani, D.~de~Florian, M.~Grazzini, and P.~Nason, ``{Soft
  gluon resummation for Higgs boson production at hadron colliders}'',}
  \textit{ JHEP} \textbf{ 07} (2003) 028,
  \href{http://dx.doi.org/10.1088/1126-6708/2003/07/028}{\doi{10.1088/1126-6708/2003/07/028}},
  \href{http://www.arXiv.org/abs/hep-ph/0306211}{\texttt{arXiv:hep-ph/0306211}}.

\bibitem{Cacciari:2003fi}
M.~Cacciari\hrefCMSnoop {}{ {et~al.}, ``{The \ttbar cross-section at 1.8 TeV
  and 1.96 TeV: A study of the systematics due to parton densities and scale
  dependence}'',} \textit{ JHEP} \textbf{ 04} (2004) 068,
  \href{http://dx.doi.org/10.1088/1126-6708/2004/04/068}{\doi{10.1088/1126-6708/2004/04/068}},
  \href{http://www.arXiv.org/abs/hep-ph/0303085}{\texttt{arXiv:hep-ph/0303085}}.

\bibitem{Kalogeropoulos:2018cke}
\hrefCMSnoop {}{A.~Kalogeropoulos and J.~Alwall, ``{The SysCalc code: A tool to
  derive theoretical systematic uncertainties}'',} 2018.
  \href{http://www.arXiv.org/abs/1801.08401}{\texttt{arXiv:1801.08401}}.

\bibitem{CMS-PAS-LUM-17-001}
\href {https://cds.cern.ch/record/2257069}{{CMS Collaboration}, ``{CMS
  luminosity measurements for the 2016 data-taking period}'',} {CMS Physics
  Analysis Summary,} CMS-PAS-LUM-17-001, 2017.

\bibitem{CMS-PAS-LUM-17-004}
\href {https://cds.cern.ch/record/2621960}{{CMS Collaboration}, ``{CMS
  luminosity measurement for the 2017 data-taking period at $\sqrt{s} =
  13~\mathrm{TeV}$}'',} {CMS Physics Analysis Summary,} CMS-PAS-LUM-17-004,
  2018.

\bibitem{CMS-PAS-LUM-18-002}
\href {https://cds.cern.ch/record/2676164}{{CMS Collaboration}, ``{CMS
  luminosity measurement for the 2018 data-taking period at $\sqrt{s} =
  13~\mathrm{TeV}$}'',} {CMS Physics Analysis Summary,} CMS-PAS-LUM-18-002,
  2019.

\bibitem{Junk_1999}
\hrefCMSnoop {}{T.~Junk, ``Confidence level computation for combining searches
  with small statistics'',} \textit{ Nucl. Instrum. Meth. A} \textbf{ 434}
  (1999) 435,
  \href{http://dx.doi.org/10.1016/s0168-9002(99)00498-2}{\doi{10.1016/s0168-9002(99)00498-2}}.

\bibitem{Read_2002}
\hrefCMSnoop {}{A.~L. Read, ``Presentation of search results: {the $\CL_{s}$
  technique}'',} \textit{ J. Phys. G} \textbf{ 28} (2002) 2693,
  \href{http://dx.doi.org/10.1088/0954-3899/28/10/313}{\doi{10.1088/0954-3899/28/10/313}}.

\bibitem{Cowan:2010js}
\hrefCMSnoop {}{G.~Cowan, K.~Cranmer, E.~Gross, and O.~Vitells, ``Asymptotic
  formulae for likelihood-based tests of new physics'',} \textit{ Eur. Phys. J.
  C} \textbf{ 71} (2011) 1554,
  \href{http://dx.doi.org/10.1140/epjc/s10052-011-1554-0}{\doi{10.1140/epjc/s10052-011-1554-0}},
  \href{http://www.arXiv.org/abs/1007.1727}{\texttt{arXiv:1007.1727}}.
[Erratum: \DOI{10.1140/epjc/s10052-013-2501-z}].
%%CITATION = ARXIV:1007.1727;%%.

\end{thebibliography}\endgroup
